\documentclass[english]{iopart}

\usepackage{graphicx}
\usepackage{url}
\newcounter{fig}
\begin{document}

\title[Renormalization, Hauptmoduls]
{\Large Renormalization, isogenies and rational 
symmetries of differential equations}
\vskip .3cm 
{\em Submitted November 29th 2009. Revised: 20 January 2010 } 

\author{A. Bostan$^\P$,
S. Boukraa$^\dag$, S. Hassani$^\S$, 
J.-M. Maillard$^{||}$,\\
J-A. Weil$^\pounds$, N. Zenine$^\S$ and N. Abarenkova$^\pounds$}
\address{$^\P$ \ INRIA Paris-Rocquencourt, 
Domaine de Voluceau, B.P. 105
78153 Le Chesnay Cedex, France} 
\address{\dag LPTHIRM and D\'epartement d'A{\'e}ronautique,
 Universit\'e de Blida, Algeria}
\address{\S  Centre de Recherche Nucl\'eaire d'Alger, 
2 Bd. Frantz Fanon, BP 399, 16000 Alger, Algeria}
\address{$||$ LPTMC, UMR 7600 CNRS, 
Universit\'e de Paris, Tour 24,
 4\`eme \'etage, case 121, 
 4 Place Jussieu, 75252 Paris Cedex 05, France} 
\address{$^\pounds$ \  XLIM, Universit\'e de Limoges, 
123 avenue Albert Thomas,
87060 Limoges Cedex, France} 
\address{$^\pounds$ \   St Petersburg Department of 
Steklov Institute of Mathematics,
27 Fontanka, 191023 St. Petersburg,  Russia} 
 
\ead{alin.bostan@inria.fr, maillard@lptmc.jussieu.fr,
 maillard@lptl.jussieu.fr, 
 njzenine@yahoo.com, boukraa@mail.univ-blida.dz,
 jacques-arthur.weil@unilim.fr,  nina@pdmi.ras.ru}

\begin{abstract}
We give an example of infinite order rational 
transformation that leaves a linear differential 
equation covariant. This example can be seen as a 
non-trivial but still simple illustration
of an exact representation of the renormalization group.

\end{abstract}

\vskip .5cm


\noindent {\bf AMS Classification scheme numbers}: 34M55, 
47E05, 81Qxx, 32G34, 34Lxx, 34Mxx, 14Kxx 


\section{Introduction}
\label{int}
\vskip .1cm

 There is no need to underline 
 the success of the renormalization group revisited
 by Wilson~\cite{Migdal,Fisher} 
 which is nowadays seen as a fundamental
 symmetry in lattice statistical mechanics or field theory.  
 It contributed to promote 
  2-d conformal field theories and/or scaling limits of second order phase
 transition in lattice statistical mechanics\footnote[5]{The renormalization group approach of
 important problems like first order phase 
 transitions,  commensurate-incommensurate 
 phase transitions, or off-critical
 problems being more problematic.}. If one does not 
take into account most of the 
 subtleties of the renormalization group, the simplest sketch
 of the renormalization group
corresponds to Migdal-Kadanoff decimation calculations 
 where the new coupling constants created at each step of the (real-space) 
 decimation calculations are forced\footnote[1]{In contrast
 with functional renormalization group~\cite{DSFisher,Wetter,Delam}.}
 to stay in some (slightly arbitrary) 
 finite dimensional parameter space. This drastic projection may be
 justified by the hope  that the basin of attraction of the fixed points
 of the corresponding (renormalization)  transformation in the parameter
 space is ``large enough''. 

 One heuristic 
 example is always given because it is
 one of the very few examples of {\em exact}
 renormalization, the renormalization
 of the one-dimensional Ising model without a magnetic field. It is a 
 straightforward undergraduate exercise to show that performing various
 decimations summing over every two, 
 or three or  ... $\, N$ spins, one gets 
{\em  exact generators of the renormalization group} reading 
 $\,T_N: \, \, t  \, \, \, \rightarrow \, \, \, t^N$ 
where $t$ is (with standard  notations) the high temperature variable
 $ \, t \, = \, \, \tanh(K)$.
It is easy to see that these transformations $\,T_N$, depending on the
 integer $\, N$, commute together. Such an {\em exact symmetry}
 is associated with a covariance of the  partition function per site 
$\, Z(t)\, = \, \, C(t) \cdot Z(t^2)$. In this particular case 
one recovers the (very simple) expression of 
 the  partition function per site,
 $\, \, 2 \, \cosh(K)$, as an infinite product 
 of the action of (for instance) $\,T_2$ on 
the cofactor $\, C(t)$. In this
 very simple case, this corresponds to using the 
identity (valid for $\, |x| <1$):
\begin{eqnarray}
 \prod_{n=0}^{\infty} \, \Bigl(1+x^{2^n}\Bigr) 
\,\, = \, \, \,\, {{1} \over {1-x}}.
\nonumber   
\end{eqnarray}
For $\,T_3: \, \, t  \, \, \, \rightarrow \, \, \, t^3$
 one must use the identity
\begin{eqnarray}
 \prod_{n=0}^{\infty} \, \Bigl(1\, +x^{3^n}\, +x^{2\cdot 3^n}\Bigr) 
\,\,  = \, \, \,\, \, \,
\prod_{n=0}^{\infty} \, 
\Bigl({{1\, -x^{3^{n+1} }} \over {1\, -x^{3^n}  }} \Bigr)
 \,\,  = \, \,\ \,\, \,{{1} \over {1-x}}.
\nonumber   
\end{eqnarray}
and for $\,T_N: \, \, t  \, \, \, \rightarrow \, \, \, t^N$ 
a similar identity where the $\, 3$ in the exponents
is changed into $\, N$.

Another simple heuristic example is
the one-dimensional Ising model {\em with a magnetic field}. 
Straightforward calculations enable to get  an 
infinite number of  exact generators of the corresponding 
renormalization group, represented as {\em rational}
 transformations\footnote[2]{One simply verifies that these 
transformations reduce to the previous
$\,T_N: \, \, t  \, \, \, \rightarrow \, \, \, t^N$ 
in the $z\, = \, 1$
limit (no magnetic field).}:
\begin{eqnarray}
\label{TN}
T_N: \quad (x, \, z) \qquad  \longrightarrow \qquad 
T_N(x, \, z)\, = \, \, (x_N, \, z_N)
\end{eqnarray}
where the first two transformations $\,T_2$ and $\,T_3$
read in terms of the two (low-temperature well-suited and 
fugacity-like) variables
$\, x \, = \, e^{4\, K}$ and $\, z \, = \, e^{2\, H}$:
\begin{eqnarray}
&&x_2 \, = \, \, \, 
{\frac { \left( x+z \right) 
 \left( 1+xz \right) }{x \cdot (1+z)^{2}}}, 
 \qquad \quad z_2 \, = \, \, \,  
 z \cdot {\frac {(1\, +xz) }{x+z}}, 
   \nonumber
\end{eqnarray}
and: 
\begin{eqnarray}
&&x_3 \, = \, \, \, x \cdot {\frac {\left( {z}^{2}x+2\,z+1 \right) 
 \left( {z}^{2}+2\,z+x \right) }
{ \left( {z}^{2}x+z+xz+x \right)^{2}}}, \,\,\, \,\, \, 
 z_3 \, = \, \, \, z \cdot
{\frac { z^{2}\, x\, +2\,z\, +1}{{z}^{2}\, +2\,z\, +x}}.   
   \nonumber
\end{eqnarray}
One simply verifies that these rational 
transformations of two (complex) variables commute.
This can be checked by formal calculations 
for  $\,T_N$ and $\,T_M$ for any $\, N$ and $M$ less than $30$,
and one can easily verify a fundamental
 property expected for renormalization
group generators: 
\begin{eqnarray}
\label{TMN}
 T_N \cdot T_M\,\, = \,\,\,\,  T_M \cdot T_N\,\, = \,\, \,\, T_{NM}, 
\end{eqnarray}
where the ``dot'' denotes the {\em composition}
 of two transformations. 
The infinite number of these rational transformations of two (complex)
variables (\ref{TN}) are thus a {\em rational representation of the
positive integers together with their product}.   
Such rational transformations can be 
studied ``per se'' as discrete dynamical
systems, the iteration of any of these various
 exact generators corresponding to
an orbit of the renormalization group.  

Of course these two examples of exact representation 
of the renormalization group are extremely degenerate
 since they correspond to one-dimensional
 models\footnote[5]{For instance the fixed points of
(\ref{TN}) are not isolated fixed points but lie on (an infinite number)
of genus zero curves ...}. Migdal-Kadanoff decimation will quite
systematically  yield {\em rational}\footnote[9]{In well-suited
 Boltzmann weight variables like $x$ and $z$ in (\ref{TN}), and
not in (bad) variables like $\, K$, the coupling constants or 
the temperature.} transformations similar to (\ref{TN})
in two, or more, variables\footnote[4]{Such  representations 
of the renormalization group
are not {\em exact} representations 
 (the exact transformation acts in an infinite number of parameters)
but some authors tried to define ``improved'' 
renormalization transformations
imposing the compatibility (commutation) of the 
renormalization transformations with some known exact symmetries of
 the model (Kramers-Wannier duality, gauge symmetries ...).}. 
Consequently, they are never (except ``academical'' self-similar models)
exact representations of the renormalization group.
 The purpose of this paper is to provide simple (but {\em non trivial})
 examples of {\em exact} renormalization 
transformations that are not degenerate 
 like the previous  transformations on one-dimensional
 models\footnote[3]{For which the partition function
 or other physical quantities are algebraic functions.}.
 In several papers~\cite{broglie,bo-ha-ma-ze-07b}
 for Yang-Baxter integrable models
 with a canonical genus one 
parametrization~\cite{Automorphisms,Baxterization} (elliptic functions
 of modulus $\, k$), we underlined that the {\em exact}
generators of the renormalization group must necessarily identify
 with the various isogenies which amounts to multiplying or 
dividing $\tau$, the ratio of the two periods of the elliptic curves, 
by an integer.  The simplest example is the {\em Landen
 transformation}~\cite{bo-ha-ma-ze-07b} 
which corresponds to multiplying ({\em or dividing} because of the 
modular group symmetry $\tau \,\leftrightarrow \, 1/\tau$),
 the ratio of the two periods:  
\begin{eqnarray} 
\label{Landen}
k \, \quad  \longrightarrow \, \quad k_L \, = \, \, 
{{2 \sqrt{k}} \over {1+k}},  \qquad \qquad 
  \tau \, \, \longleftrightarrow \, \, 2\, \tau.
\end{eqnarray} 
The other transformations\footnote[1]{See 
for instance (2.18) in~\cite{Canada}.} 
correspond to 
$\tau \,\leftrightarrow \, N \cdot \tau$,
for various integers $\, N$. In the (transcendental)
variable $\tau$,
 it is clear that they satisfy
relations like (\ref{TMN}). However, in 
the natural variables of the model
 ($e^{K}, \, \tanh(K), \, k\, = \, \sinh(2\, K)$, not transcendental
variables like $\tau$), these transformations
 are {\em algebraic} transformations
corresponding  in fact to the {\em fundamental modular curves}. 
For instance (\ref{Landen}) corresponds to the 
{\em genus zero fundamental modular curve}
\begin{eqnarray} 
\label{fundmodular1}
&&j^2 \cdot j'^2 -(j+j') \cdot
 (j^2+1487 \cdot j\, j' \, +j'^2) \nonumber \\
&& \qquad  \,  
 +3 \cdot 15^3 \cdot (16\, j^2\, -4027\, j\, j' \, +16\, j'^2) \\
 &&   \qquad \,  -12\cdot 30^6\cdot (j+j') \, +8\cdot 30^9
\, = \, \, \, \, 0 , \nonumber
\end{eqnarray} 
or:
\begin{eqnarray} 
\label{fundmodular}
&&5^9\, v^3\, u^3\, -12 \cdot 5^6 \, u^2\, v^2 \cdot (u+v)\, 
+375\, \, u\, v \cdot  (16\, u^2\, +16\, v^2\, -4027\, v\, u)
 \nonumber \\
&&\quad \quad -64\, (v+u)\cdot (v^2+1487\, v\, u\, +u^2) \, \, 
+2^{12}\cdot  3^3 \cdot  u\, v  \, = \, \, 0,
\end{eqnarray} 
which relates the two Hauptmoduls $\, u\, = \, \, 12^{3}/j(k)$,
 $\,\,\, v\, = \, \, 12^{3}/j(k_L)$:
\begin{eqnarray} 
&&j(k) \, = \, \, \, \, 256
\cdot {{(1-k^2+k^4)^3} \over {k^4 \cdot (1-k^2)^2}},
 \, \, \nonumber \\
&&\qquad \qquad  j(k_L) \, = \, \, \, \, 
16 \cdot {\frac { (1+14\,{k}^{2}+{k}^{4})^3}{
 (1-{k}^{2})^{4} \cdot {k}^{2} }}.
 \nonumber 
\end{eqnarray}
 One verifies easily that (\ref{fundmodular1}) 
is verified with $j\, = \, \, j(k)$ and $j'\, = \, \, j(k_L)$.

The selected values of $\, k$,
 the modulus of elliptic functions,  $\, k\, = 0, \, 1$ 
are actually {\em fixed points of the Landen transformations}. 
The Kramers-Wannier duality $k\, \leftrightarrow \, 1/k$
maps $\, k\, = 0$ onto  $\, k\, = \, \infty$. 
For the Ising (resp. Baxter) model these selected 
values of $\, k$ correspond to the three
selected subcases of the model  ($\, T \, = \, \infty$, $\, T \, = \, 0$ 
and the critical temperature $\, T \, = \, T_c$), for which
the elliptic parametrization of the model
degenerates into a rational parametrization~\cite{bo-ha-ma-ze-07b}. 
We have the same property for all the other 
algebraic modular curves corresponding 
to $\, \tau \, \leftrightarrow \, N \cdot \tau$.
This is certainly the main property most physicists expect for 
an exact representation of a {\em generator of the renormalization
group}, namely that it maps a generic point of the parameter space onto 
 the critical manifold (fixed points).
 Modular transformations are, in fact,
 the only transformations to be
compatible with all the other symmetries of the Ising (resp. Baxter) 
model like for instance, the gauge transformations, some extended 
$\, sl(2) \times  sl(2) \times sl(2) \times  sl(2)$ 
symmetry~\cite{BeMaVi92}, etc.  It has also been
 underlined in~\cite{broglie,bo-ha-ma-ze-07b}
that seeing (\ref{Landen}) as a transformation on {\em complex variables}
(instead of real variables) 
provides two other complex fixed points which actually correspond to
{\em complex multiplication} for the elliptic curve, and are, actually,  
fundamental new singularities\footnote[4]{Suggesting an 
understanding~\cite{bo-ha-ma-ze-07b,bo-ha-ma-ze-07}
 of the quite rich structure
of infinite number of the singularities of the $\, \chi^{(n)}$'s 
in the complex plane from a Hauptmodul 
approach~\cite{bo-ha-ma-ze-07b,bo-ha-ma-ze-07}.
Furthermore the notion of {\em Heegner numbers}
 is closely linked to the isogenies
mentioned here~\cite{bo-ha-ma-ze-07b}. An exact 
value of the $j$-function
$\, j(\tau)$ corresponding one of the first Heegner number is,
for instance,  
$\, j(1\, +i)\, = \, \,12^3$.}
discovered on the  $\, \chi^{(3)}$ linear 
ODE~\cite{ze-bo-ha-ma-04,ze-bo-ha-ma-05,ze-bo-ha-ma-05c}.
In general, this underlines the deep relation between 
the renormalization group and the theory 
of elliptic curves in a deep sense,
namely {\em isogenies of elliptic curves,
 Hauptmoduls}\footnote[1]{It 
should be recalled that the mirror
symmetry found with Calabi-Yau 
manifolds~\cite{Candelas,LianYau,Doran,Doran2,Kratten}
 can be seen as higher order
generalizations of Hauptmoduls. We thus have already generalizations
of this identification of the renormalization and modular structure
when one is not restricted to elliptic curves anymore.},
 {\em modular curves and modular forms}.

Note that an algebraic transformation like (\ref{Landen}) or 
(\ref{fundmodular}) cannot be obtained from 
any {\em local} Migdal-Kadanoff transformation which naturally 
yields {\em rational} transformations: an exact renormalization
group transformation like (\ref{Landen}) can only be 
deduced from {\em non-local} decimations. The emergence of modular
transformations as representations of exact generators of 
the renormalization group explains, in a quite subtle way,
the difficult problem of how  renormalization
group transformations can be compatible 
with {\em reversibility}\footnote[2]{The 
fact that the renormalization group
must be reversible has apparently been totally forgotten 
by most of the authors who just see
a semi-group corresponding to forward 
iterations converging to the critical
points (resp. manifolds).}
(iteration forward and backwards). An
 algebraic modular transformation (\ref{fundmodular})
corresponds to $\tau \, \rightarrow \, 2 \, \tau$
{\em and} $\tau \, \rightarrow \, \, \tau/2$
{\em in the same time, as a consequence of the modular
group symmetry} $\tau \, \leftrightarrow \,  \, 1/\tau$.

A simple rational parametrization\footnote[9]{Corresponding to
Atkin-Lehner polynomials and Weber's functions.} of the genus zero
 modular curve (\ref{fundmodular}) reads:
\begin{eqnarray}
\label{para}
u \, = \, \, 1728\,{\frac {z}{ \left( z+16 \right)^{3}}}, \qquad  
v\, = \, \, 1728\,\,{\frac {{z}^{2}}{ \left( z+256 \right)^{3}}}
\, = \, \, u\Bigl( {{2^{12}} \over {z}}\Bigr).
\end{eqnarray}
Note that the previously mentioned reversibility
is also associated with the fact that 
the modular curve (\ref{fundmodular})
is invariant by $\, u \, \leftrightarrow \, v$, 
and, within the previous rational parametrization (\ref{para}), 
with the fact that permuting  $\, u$ and  $\, v$
corresponds\footnote[3]{Conversely, and more precisely,
writing $\, 1728 \,z^2/(z+256)^3\,  = \,\, \,1728\, z'/(z'+16)^3$
 gives the
 Atkin-Lehner~\cite{Atkin} involution $\,z \cdot z'= \,2^{12}$, 
together with the quadratic relation
 $\, z^² \, - \, z \, z'^² \, -48 \, z \, z' \, -4096 \, z'
\, = \, \, 0$. }
 to the Atkin-Lehner involution
 $\, z  \, \leftrightarrow \,2^{12}/z$.

 For many Yang-Baxter integrable 
models of lattice statistical mechanics
 the physical quantities (partition function per site,
correlation functions, ...) are solutions of selected\footnote[5]{They
are not only Fuchsian, the corresponding linear differential operators
are globally nilpotent or $\, G$-operators~\cite{bo-bo-ha-ma-we-ze-09}.} 
linear differential equations.
For instance the partition function 
per site of the square (resp. triangular, etc.)
 Ising model is an integral of an elliptic integral of the third kind.
It would be too complicated to show the precise covariance of these
physical quantities with respect to (algebraic) modular transformations
like (\ref{fundmodular}). 
Instead, let us give, here, an illustration 
of the non-trivial action of the 
renormalization group on some elliptic function that actually occurs
in the 2-D Ising model: a weight-one modular form. 
This modular form actually, and remarkably, 
emerged~\cite{bo-bo-ha-ma-we-ze-09} in a second
order linear differential operator factor denoted $\, Z_2$
occurring~\cite{ze-bo-ha-ma-04} for $\, \chi^{(3)}$,
  and that the reader can think as
a physical quantity solution of a particular linear ODE replacing the too 
 complicated integral of an elliptic integral of the third kind.
Let us consider the second order linear differential operator
($D_z$ denotes $\, d/dz$):
\begin{eqnarray}
\alpha \, = \,\,\, \, \,\, \, D_z^{2} \,\, \, 
+{\frac { \left( {z}^{2}+56\,z+1024 \right)}{ z \cdot (z+16)\,  (z+64)  }} 
\cdot D_z\,\, \,
- \,{\frac {240}{ z \cdot (z+16)^2  \,(z+64) }},
\nonumber 
\end{eqnarray}
which has the (modular form) solution:
\begin{eqnarray}
\label{cov}
&&_2F_ 1 \Bigl([1/12,5/12],[1];  
 1728\,{\frac {z}{ (z+16)^3}}\Bigr)
 \,\,\,   \\
&& \quad \quad \, = \, \, \,\, 
2 \cdot \Bigl({\frac {z+256}{z+16}}\Bigr)^{-1/4}
\cdot \, \, \, 
_2F_ 1 \Bigl([1/12,5/12],[1];
  1728\,\,{\frac {{z}^2}{ (z+256)^3}}\Bigr).
 \nonumber 
\end{eqnarray}
Do note that the two pull-backs in the arguments of the 
{\em same} hypergeometric function are {\em actually related
by the modular curve relation}  (\ref{fundmodular}) (see (\ref{para})).
The covariance  (\ref{cov}) is thus the very expression of a
modular form property with respect to a modular
transformation ($\tau \, \leftrightarrow \, 2 \, \tau$)
corresponding to the modular transformation  (\ref{fundmodular}).

The hypergeometric function at the rhs of (\ref{cov})
is solution of the second order linear differential operator 
\begin{eqnarray}
\beta  \, = \, \, \, \, D_z^{2} \,\,  
+ {\frac {{z}^{2}+416\,z+16384}{ (z+256)\,  
 \, (z+64)\,  z }}
\cdot D_z\, \, 
-\,{\frac {60}{ (z+64)  \, (z+256)^{2}}},
\nonumber 
\end{eqnarray}
which is the transformed of operator $\, \alpha$ 
by the Atkin-Lehner duality
$\, z \, \leftrightarrow \, \, 2^{12}/z$,  and, 
 also, a conjugation of $\, \alpha$:
\begin{eqnarray}
\beta \, = \, \,  \, 
\Bigl({\frac {z+16}{z+256}}\Bigr)^{-1/4} \cdot \alpha  \cdot 
\Bigl({\frac {z+16}{z+256}}\Bigr)^{1/4}.
\end{eqnarray}

Along this line we can also recall that the
 (modular form) function\footnote[2]{Where
$\, j$ is typically the $\, j$-function~\cite{Canada,j}.}:
\begin{eqnarray}
 F(j) \, = \, \,\, \,\,\, 
j^{-1/12} \cdot \,
 _ 2F_ 1 \Bigl([1/12,5/12],[1];   {{12^3} \over {j}} \Bigr), 
\end{eqnarray}
verifies:
\begin{eqnarray}
 F\Bigl( {\frac { \left( z+16 \right)^{3}}{z}}  \Bigr) 
\,\, = \, \, \,\,  
2 \cdot  z^{-1/12}
\cdot F\Bigl( {\frac { \left( z+256 \right) ^{3}}{{z}^{2}}}  \Bigr). 
\end{eqnarray}

A relation like (\ref{cov})
 is a straight generalization 
of the covariance we had in the one-dimensional
model $\, Z(t)\, = \, \, C(t) \cdot Z(t^2)$, 
which basically amounts to seeing
the partition function
per site as some ``automorphic function'' with respect
 to the renormalization group, the simple renormalization group
transformation $t \, \rightarrow \, t^2$ being replaced by 
the algebraic modular transformation (\ref{fundmodular})
corresponding to $\tau \, \leftrightarrow \, 2 \, \tau$
(that is the Landen transformation (\ref{Landen})). 
 
 We have here all the ingredients 
for seeing the identification of exact algebraic representations
of the renormalization group with the modular curves structures
we tried so many times to promote (preaching in the desert)
 in various papers~\cite{broglie,bo-ha-ma-ze-07b}. 
However, even if there are no difficulties, just subtleties, these
Ising-Baxter examples of exact algebraic representations
of the renormalization group already require some serious knowledge
of the modular curves, modular forms and Hauptmoduls
in the theory of elliptic curves, mixed with the subtleties
 naturally associated with the various
 branches of such algebraic  (multivalued) transformations. 

The purpose of this paper is to present another 
elliptic hypergeometric function and other
 much simpler (Gauss hypergeometric) 
second order linear differential operators  
covariant by infinite order rational transformations. 

The replacement of {\em algebraic (modular) transformations} by simple 
{\em rational} transformations will enable us to display
a complete {\em explicit description of an exact representation of the 
renormalization group} that any graduate student can completely dominate.

\section{Infinite number of rational symmetries 
 on a Gauss hypergeometric ODE}
\label{infinitegauss}
\vskip .1cm
Keeping in mind modular form expressions like (\ref{cov}), 
let us recall a particular Gauss
hypergeometric function
introduced by  R. Vidunas in~\cite{Vidunas}:
\begin{eqnarray}
\label{vid}
&&_2F_1\Bigl( [{{1} \over {2}}, {{1} \over {4}}],
 [{{5} \over {4}}];\,  z\Bigr) 
\, = \, \,  \, 
{{1} \over {4}} \cdot z^{-1/4} \cdot
 \int_0^{z}\, t^{-3/4} \, (1-t)^{-1/2} dt 
 \, \nonumber \\
&& \qquad \quad \quad  \,  = \, \, 
(1-z)^{-1/2} \cdot \, \, \, 
_2F_1\Bigl( [{{1} \over {2}}, {{1} \over {4}}], [{{5} \over {4}}];
 \, {{-4\, z} \over {(1-z)^2}} \Bigr). 
\end{eqnarray}
This hypergeometric function corresponds to the integral of
 a holomorphic form on a {\em genus-one} curve 
$\, P(y, \, t) \, = \, 0$:
\begin{eqnarray}
\label{ellip}
{{dt} \over {y}}, \qquad \hbox{with:} \qquad \qquad 
 y^4 \, - \, \, t^3 \cdot (1-t)^2 \, \, = \, \, \, 0.
\end{eqnarray}
Note that the function
\begin{eqnarray}
\label{defF}
{\cal F}(z)\,\, = \, \,\,\, \,
 z^{1/4} \cdot  \, \,
 _2F_1\Bigl( [{{1} \over {2}}, {{1} \over {4}}],
 [{{5} \over {4}}];\,  z\Bigr), 
\end{eqnarray}
which is exactly an integral of an algebraic function,  
has an extremely simple covariance property
 with respect to the {\em infinite order
rational} transformation $\, z \, \rightarrow \, -4\, z/(1-z)^2$:
\begin{eqnarray}
\label{F}
{\cal F}\Bigl({{-4\, z} \over {(1-z)^2}} \Bigr)
\,\, \,  = \, \, \,  \,\,\, (-4)^{1/4} \cdot {\cal F}(z). 
 \end{eqnarray}
The occurrence of this specific infinite order transformation is
reminiscent of Kummer's quadratic relation
\begin{eqnarray}
\label{Kummer}
&&_2F_1\Bigl( [a, \, b], [1+a-b];\,  z\Bigr)\, =\, \,\\
&& \qquad \, =\, \,(1-z)^{-a} \cdot \, \,
 _2F_1 \left( [\frac{a}{2}, \frac{1+a}{2} -b], [1+a-b];
 \,  -\frac{4\, z}{(1-z)^2} \right), 
\nonumber 
\end{eqnarray}
but it is crucial to note that, relation (\ref{F})
does not relate two different functions, 
but is an ``automorphy'' relation
on the {\em same function}. 

It is clear from the previous paragraph 
that we want to see such functions
as 'ideal' examples of physical functions
 covariant by an exact (here, rational)
generator of the renormalization group. The function
 (\ref{defF}) is actually solution of the 
second order linear differential operator:
\begin{eqnarray}
\label{Omega}
&&\Omega \, = \, \, \, \, \,
  D_{z}^2 \,\, 
+  {{1} \over {4}} \, {{3 \, -5\,z} \over {z \cdot (1-z)}}  \cdot D_z 
\,\,\, = \,\,\, \,\,  \omega_1 \cdot  D_z, 
\qquad  \qquad \hbox{with:}  \\
&& \omega_1  \, = \, \,\, \,\,
 D_{z} \,\, +  {{1} \over {4}} \, {{3 \, -5\,z} \over {z \cdot (1-z)}} 
 \, \,  = \, \,\, \,\,
 D_z \, +\, \, \, 
 {{1} \over {4}} \cdot 
 {{ d\ln (z^3 \, (1-z)^2)} \over {dz}}.\nonumber
\end{eqnarray}
From the previous expression of 
$\, \omega_1$ involving a log-derivative of a rational function
it is obvious that this second order linear differential operator
has two solutions, the constant function
 and an integral of an algebraic function.
Since these two solutions behave very simply under the  infinite order
rational transformation $\, z \, \rightarrow \, -4\, z/(1-z)^2$, it is
 totally and utterly natural to see how the linear differential
operator $\, \Omega$ transforms under the rational change of variable
$\, z \, \rightarrow \, R(z) \, = \, \,  -4\, z/(1-z)^2$
(which amounts to seeing how the two order-one operators
$\, \omega_1$ and $\,  D_z$ transform). It 
is a straightforward calculation
 to see that introducing the cofactor $\, C(z)$
which is the inverse of the derivative of $\, R(z)$ 
\begin{eqnarray}
C(z) \, = \,  \,\, \, 
-\, {{1} \over {4}} \cdot {\frac { (1-z)^{3}}{1+z}},
 \qquad \qquad  
{{1} \over {C(z)}}   \, = \,  \, \, {{ d R(z)} \over {dz}},
\end{eqnarray}
$\,  D_z$ and $\, \omega_1$ respectively 
transform under the rational change of variable
$\, z \, \rightarrow \, R(z) \, = \, \,  -4\, z/(1-z)^2$
 as:
\begin{eqnarray}
\label{omega1}
&&D_z \, \longrightarrow \, \, \, C(z) \cdot D_z, \,  \,  \quad 
\omega_1 \,  \longrightarrow \, \, \, (\omega_1)^{(R)}  \, = \, \, \, 
 C(z)^2  \cdot \omega_1 \cdot {{1} \over {C(z)}}, \\
\label{omega1bis}
&&\hbox{yielding:} \qquad  \qquad
 \Omega \, \, \longrightarrow \, \, \, C(z)^2  \cdot \Omega. 
\end{eqnarray}
Since  $\, z \, \rightarrow \, -4\, z/(1-z)^2$ is of infinite order,
the second order linear differential operator (\ref{Omega})  has 
{\em an infinite number of rational symmetries} (isogenies):
\begin{eqnarray}
\label{iterR}
&&z \quad \longrightarrow \quad {{-4\, z} \over {(1-z)^2}}
 \quad \longrightarrow \quad 
16\cdot {\frac { (1-z)^{2} \cdot z}{ (1+z)^{4}}}
\quad \longrightarrow    \nonumber \\
&& \qquad \longrightarrow \quad
 -64\cdot {\frac { (1-z)^{2}  \, (1+z)^{4} \, z}
{ \, (1-6\,z\, +z^2)^4}}
\quad \longrightarrow \quad \cdots 
\end{eqnarray}

Once we have found a second order linear differential 
operator (written in a unitary
or monic form) $\Omega$, covariant by
the infinite order rational transformation
 $\, z \, \rightarrow \, -4\, z/(1-z)^2$,
it is natural to seek for higher order 
linear differential operators also covariant by
 $\, z \, \rightarrow \, -4\, z/(1-z)^2$. One easily verifies that 
the successive symmetric powers of  $\Omega$ 
are (of course ...) also covariant. The symmetric
 square of $\, \Omega$,
\begin{eqnarray}
\label{vic2}
D_z^3 \,\,\, 
  + \,   {{3} \over {4}}\,{\frac {3-5\,z}{ (1-z)\, z }}
 \cdot  D_z^2
\,\, \,  +\, {{3} \over {8}}\, {\frac {1-5\,z}{ (1-z)\, z^2 }}
\cdot  D_z,
\end{eqnarray}
factorizes in simple order-one operators: 
\begin{eqnarray}
\Bigl(D_z \, +
 \,   {{2} \over {4}} \,{\frac {3-5\,z}{ (1-z)\, z }} \Bigr) 
 \cdot
 \Bigl(D_z \, +
 \,  {{1} \over {4}} \,{\frac {3-5\,z}{ (1-z)\, z}} \Bigr) 
 \cdot  D_z,
\end{eqnarray}
and, more generally, the symmetric $\, N$-th power\footnote[1]{Such formula is actually valid
for $\,\Omega_A \, = \, \, (D_z \, +A(z)) \cdot D_z$ for any $\, A(z)$. Denoting
 ${\cal S}_N$ symmetric $\, N$-th power 
of  $\,\Omega_A$  one has 
 ${\cal S}_N\, = \, \, (D_z \, +A(z)) \cdot {\cal S}_{N-1}$.} of $\, \Omega$
reads
\begin{eqnarray}
&&\Bigl(D_z \, + \,   {{N} \over {4}}
 \,{\frac {3-5\,z}{z \left( 1-z \right) }} \Bigr)  \cdot
\Bigl(D_z \, + \,   {{N-1} \over {4}}
 \,{\frac {3-5\,z}{z \left( 1-z \right) }} \Bigr) 
 \, \,\cdots  \\
&& \qquad \qquad \quad \cdots  \, \,
\Bigl(D_z \, + \,  {{1} \over {4}} 
\,{\frac {3-5\,z}{z \left( 1-z \right) }} \Bigr) 
 \cdot  D_z.  \nonumber 
\end{eqnarray}
The covariance of such expressions is the
 straight consequence of the fact that
the order-one factors
\begin{eqnarray}
\omega_k \, = \, \, \, D_z \,\, 
  + \,   {{k} \over {4}} \,{\frac {3-5\,z}{z \cdot (1-z) }},
\qquad \quad k\, = \, \, 0, \, \, 1, \, \, \cdots, \, \,  N, 
\end{eqnarray}
transform very simply under 
 $\, z \, \rightarrow \, -4\, z/(1-z)^2$:
\begin{eqnarray}
\label{omegaktrs}
\omega_k  \quad \longrightarrow \quad (\omega_k)^{(R)} 
\,\, = \,\,\, \, \, \Bigl(C(z)\Bigr)^{k+1} \cdot 
\omega_k 
 \cdot \Bigl(C(z)\Bigr)^{-k}. 
\end{eqnarray}
More generally, let us consider a rational transformation 
$\, z \, \rightarrow \, R(z)$, the corresponding cofactor 
$\, C(z) \, = \, \,  1/R'(z)$, and the order-one operator 
$\, \omega_1 \, = \, \, D_z \, \, + \, A(z)$.
We have the identity:
\begin{eqnarray}
\label{ident}
C(z) \cdot D_ z \cdot \bigl({{1} \over {C(z)}}\Bigr) 
\, \,  = \,\,    \, \, \, \, 
 D_ z \, \,  -\, {{d \ln(C(z))} \over {dz}}.
\end{eqnarray}
The change of variable $\, z \, \rightarrow \, R(z)$
on  $\, \omega_1$  reads:
\begin{eqnarray}
D_z \, + \, \, A(z) \quad \longrightarrow \quad   
C(z) \cdot  D_z \, + \, \, A(R(z))  \, = \, \, \, \,
C(z) \cdot \Bigl( D_z \, + \, \, B(z) \Bigr). 
\, \nonumber
\end{eqnarray}
We want to impose that this rhs expression can be written 
(see (\ref{omega1})) as:
\begin{eqnarray}
C(z)^2 \cdot 
\Bigl(D_z \, + \, \, A(z)\Bigr) \cdot {{1} \over {C(z)}},
 \nonumber 
\end{eqnarray}
which, because of (\ref{ident}),  occurs if  
\begin{eqnarray}
B(z) \, = \,\, \,    \,\, 
A(z) \, - \, {{ d\ln(C(z))} \over {dz}},\nonumber  
\end{eqnarray}
yielding a ``Rota-Baxter-like''~\cite{Rota,Rota2}
 functional equation on $\, A(z)$ and $\, R(z)$:
\begin{eqnarray}
\label{mad}
\Bigl({{ d R(z)} \over {dz}}\Bigr)^2 \cdot A(R(z)) 
\,\,     = \,\,  \,  \,\,
    {{ d R(z)} \over {dz}} \cdot A(z) \, \, \,  
 + {{ d^2 R(z)} \over {dz^2}}. 
\end{eqnarray}

\vskip .3cm
{\bf Remark:} 
Coming back to the initial Gauss 
hypergeometric differential operator
the covariance of $\, \Omega$  becomes a conjugation.
Let us start with the Gauss hypergeometric differential operator
for (\ref{vid}):
\begin{eqnarray}
H_0\, = \, \, \, \, 8\,z \cdot (1 -z) \cdot  D_z^{2}\, \,
+ \, 2 \cdot (5\, -7\,z) \cdot  D_z\, \, -1. 
\end{eqnarray}
It is transformed by
 $\, z \, \rightarrow \, R(z) \, = \, \, -4 \, z/(1-z)^2$ 
into:
\begin{eqnarray}
&&H_1 \, = \, \, \,  \, 8\,z \cdot (1- z)\cdot  D_z^{2} \,\,
 -2\, \left( 3\,z-5 \right) \cdot  D_z \,\, +\, {{4} \over {1-z}}
  \, \nonumber \\
&&\quad \quad \quad \, = \, \, 
(1-z)^{1/2} \cdot H_0 \cdot (1-z)^{-1/2},
  \nonumber 
\end{eqnarray}
then by $\, z \, \rightarrow \, R(R(z)) \, = \, R_2(z) \,$
$ = \, \,16\, z \, (1-z)^2/(1+z)^4$ 
into:
\begin{eqnarray}
&&H_2 \, = \,\,\, \, 8\,z \cdot (1- z) \cdot  D_z^{2} \, \,
-2\,{\frac { \left( 3\,z-1 \right)  \left( z+5 \right) }{z+1}}  
 \cdot    D_z\,\,
\, +16\,{\frac {z-1}{ \left( z+1 \right)^{2}}}\, \,\nonumber \\
&&\quad \quad \quad \quad \, = \,  \, \,\, 
\Bigl({\frac {z+1}{\sqrt {z-1}}} \Bigr) \cdot H_0 
\cdot \Bigl({\frac {z+1}{\sqrt {z-1}}} \Bigr)^{-1} \nonumber 
\end{eqnarray}
and more generally for 
$\, z \, \rightarrow \,R_N \, = \,  R(R(R \cdots (R(z) \cdots )$:
\begin{eqnarray}
H_N \, = \, \,\,\, C_N \cdot H_ 0 \cdot C_N ^{-1}, \qquad 
\hbox{where:}\qquad  C_N \, = \, \,\, \,  z^{1/4} \cdot R_N^{-1/4}.
\nonumber 
\end{eqnarray}

\vskip .3cm

\subsection{A few remarks on the ``Rota-Baxter-like''  functional equation.}
\label{afewremarks}

The  functional
 equation\footnote[1]{The Rota-Baxter relation of weight 
$\Theta$ reads: $ \, R(x) \, R(y) \, 
+ \Theta R(x\, y) =\, R(R(x)\,y\, + x\,R(y))$.} (\ref{mad})
 is the (necessary and sufficient) condition for
$\, \Omega\, = (D_z\, + \, A(z)) \cdot D_z$ to be 
covariant by $\, \, z \, \rightarrow \, R(z)$.

Using the chain rule formula of derivatives of composed functions:
\begin{eqnarray}
&&{{ d\, R(R(z))} \over {dz}} \, = \, \, \,
 {{ d\, R(z)} \over {dz}} \cdot [{{ d\, R(z)} \over {dz}}(R(z))],
  \nonumber \\
&&{{ d^2\, R(R(z))} \over {dz^2}} \, = \, \,\nonumber \\
&& \quad  \quad  \quad 
{{ d^2\, R(z)} \over {dz^2}} \cdot [{{ d\, R(z)} \over {dz}}(R(z))]\, 
+ \, \Bigl({{ d\, R(z)} \over {dz}} \Bigr)^2
 \cdot [{{ d^2\, R(z)} \over {dz^2}}(R(z))], 
\nonumber 
\end{eqnarray}
one can show that, for $\, A(z)$ fixed,  the ``Rota-Baxter-like''
 functional equation (\ref{mad})
is  invariant 
by the composition of $\, R(z)$ by itself
$\, R(z) \, \longrightarrow \, \, R(R(z))$,
 $\, R(R(R(z))), \,\cdots$ 
This result can be generalized to
any composition of various $\, R(z)$'s satisfying (\ref{mad}).
This is in agreement with the fact that (\ref{mad}) 
 is the condition for
$\, \Omega\, =\,\, (D_z\, + \, A(z)) \cdot D_z$ to be 
covariant by $z \, \rightarrow \, R(z)$: it must be 
invariant by composition of $\, R(z)$'s
 (for $\, A(z)$ fixed).

Note that we have not used here the fact that
for globally nilpotent~\cite{bo-bo-ha-ma-we-ze-09} 
operators,  $\, A(z)$
and  $\, B(z)$ are necessarily log-derivatives of $\, N$-th
 roots of rational functions. For 
 $\, R(z)\, = \, \, -4\, z/(1-z)^2$:
\begin{eqnarray}
\label{glob}
&&A(z) \, = \, \, {{1} \over {4}} \cdot 
{{d \ln(a(z))} \over {dz}},
 \qquad  \quad 
B(z) \, = \, \,  {{1} \over {4}} \cdot 
{{d \ln(b(z))} \over {dz}}, 
  \\
&&a(z) \, = \, \, \, (1-z)^{2} \cdot {z}^{3}, \, \,
 \qquad \quad \quad 
b(z) \, = \, \, {z}^{3} \cdot {\frac { (1+z)^{4}}{ (1-z)^{10}}}.
\nonumber 
\end{eqnarray}
The existence of the underlying $\, a(z)$ in (\ref{glob})
consequence of a global nilpotence of the order-one 
differential operator, can however be seen in the following remark
on the zeros of the lhs and rhs terms in the functional equation 
(\ref{mad}). When $\, R(z)$ is a rational function 
(for instance $\, -4\, z/(1-z)^2$ 
or any of its iterates $\, R^{(n)}(z)$),
the lhs and rhs of (\ref{mad}) are rational expressions. The zeros are 
 roots of the numerators of these rational expressions. 
Because of (\ref{glob}) the functional equation (\ref{mad}) 
 can be rewritten (after dividing by $\, R'(z)$) as:
\begin{eqnarray}
\label{mad2}
&&\Bigl({{ d R(z)} \over {dz}}\Bigr) \cdot A(R(z)) 
\,\,  \,\,   = \,\,\,  \,\,\, \,     A(z) \, \, \,  
 + {{ d } \over {dz}} 
\Bigl(\ln\Bigl( {{ d R(z)} \over {dz}} \Bigr)\Bigr)  \\
&&\qquad \qquad \, = \, \, \, \, 
{{1} \over {4}} \cdot  {{ d } \over {dz}} \Bigl(\ln\Bigl(a(z)
 \cdot 
\Bigl({{ d R(z)} \over {dz}}\Bigr)^4 \Bigr)
\Bigr).
\nonumber 
\end{eqnarray}
One easily verifies, in our example, that
 the zeros of the rhs of (\ref{mad2}) 
come from the zeros of $\, A(R(z))$ (and not from the zeros of $\, R'(z)$
 in the lhs of (\ref{mad2})). The zeros
 of the log-derivative  rhs of (\ref{mad2})
correspond to $\, a(z) \cdot R'(z)^4 \, = \, \, \rho$, where $\,\rho$
 is a constant to be found. 
Let us consider for $\, R(z)$ the $\, n$-th iterates of $\, -4\, z/(1-z)^2$
that we denote $\, R^{(n)}(z)$.
A straightforward calculation shows that 
the zeros of $\, A(R^{(n)}(z))$ or $\, a'(R^{(n)}(z))$ (where
 $\, a'(z)$ denotes the
derivative of $a(z)$ namely $\, (z-1)\, (5\,z-3)\cdot z^2$) 
actually correspond to the general closed formula:
\begin{eqnarray}
\label{rho}
5^5 \cdot a(z) \cdot  \Bigl({{ d R^{(n)}(z)} \over {dz}}\Bigr)^4 
\,  -  \, \, 4 \cdot 3^3\cdot (-4)^n  \,\,\, = \,\,\,  \, \, 0.
\end{eqnarray}
More precisely the zeros of $\,\,  5 \cdot R^{(n)}(z) \, -3\,\,  $ 
verify (\ref{rho}), or, in other words, the numerator of 
 $\,\, \,  5 \, R^{(n)}(z) \, -3\,\,  $
divides the numerator of the lhs of (\ref{rho}).

In another case for $\, T(z)$ 
given by (\ref{Tz}), which also verifies (\ref{mad}) (see below),
the relation (\ref{rho}) is replaced by:
\begin{eqnarray}
\label{rho2}
5^5 \cdot a(z) \cdot  \Bigl({{ d T^{(n)}(z)} \over {dz}}\Bigr)^4 
\,  -  \, \, 4 \cdot 3^3 \cdot (-7\, -24\, i)^n 
 \,\,\, \,  = \,\,\,  \,  \, \, 0.
\end{eqnarray}
More generally for a rational function $\, \rho(x)$, 
obtained by an arbitrary composition
of $\, -4\, z/(1-z)^2$ and $\, T(z)$,  we would have:
\begin{eqnarray}
\label{rho3}
5^5 \cdot a(z) \cdot  \Bigl({{ d \rho(z)} \over {dz}}\Bigr)^4 
\,  -  \, \, 4 \cdot 3^3 \cdot \lambda^n 
 \,\,\, \,  = \,\,\,  \, \,  \, 0.
\end{eqnarray}
where $\, \lambda$ corresponds to: 
\begin{eqnarray}
\rho(x) \, = \, \, \,  \lambda \cdot z \, + \, \, \cdots, 
\qquad \qquad 
 \lambda \, = \, \, \Big[{{ d \rho(z)} \over {dz}}\Bigr]_{z=0}.
\end{eqnarray}

\vskip .3cm 

\subsection{Symmetries of $\, \Omega$, solutions the
 ``Rota-Baxter-like''  functional equation.}
\label{symsol}

Let us now analyse
 all the symmetries of the linear differential operator 
$\, \Omega\, =\, (D_z\, + \, A(z)) \cdot D_z$ 
by analyzing all the solutions of (\ref{mad})
for a given $\, A(z)$.
For simplicity we will restrict to 
 $\, A(z)\, = \, \, (3\, -5\,z)/z/(1-z)/4$
which corresponds to  $\, R(z) =\, -4\, z/(z-1)^2$
and all its iterates (\ref{iterR}). Let us first seek 
for other (more general) solutions that are
 {\em analytic at} $\, z \, = \, \, 0$:
\begin{eqnarray}
\label{anal}
R(z) \, = \, \, \, a_1 \cdot z \, + a_2 \cdot z^2 \, 
+ a_3 \cdot z^3 \,\,  +\, \,  \cdots  
\end{eqnarray}
It is a straightforward calculation to get,
order by order from (\ref{mad}), the successive coefficients $\, a_n$
in (\ref{anal}) as polynomial expressions (with rational 
coefficients) of the first coefficient $\, a_1$ with
\begin{eqnarray}
\label{orderbyorder}
&&a_2\, = \, \, - {{2} \over {5}} \cdot a_1  \cdot (a_1\, -1),
 \quad \quad \, \, \, \, 
a_3\, = \, \,
 {{1} \over {75}} \cdot  a_1  \cdot (a_1\, -1) \cdot (7\, a_1\, -17),
 \nonumber \\
&& a_4\, = \, \,- {{2} \over {4875}}  \cdot  a_1  \cdot (a_1\, -1) \cdot
 (41\, a_1^2\, -232\, a_1\, +366), 
\, \, \,\,\,  \,  \, \cdots \nonumber \\
&& a_n\, = \, \, -\, {{n} \over {5}}
 \cdot a_1  \cdot (a_1\, -1) \cdot {{P_n(a_1)} \over {P_n(-4) }}, 
\end{eqnarray}
where  $\, P_n(a_1)$  is a polynomial
 with integer coefficients of degree $\, n-2$. 
Since we have here a series depending on one parameter $\, a_1$
we will denote it  $\, R_{a_1}(z)$. 
This is a quite remarkable series depending
on one parameter\footnote[5]{For $\, A(z)$ given we get a 
one-parameter family of $\, R(z)$
solution of (\ref{mad}).
 Conversely, for $\, R(z)$ given 
one can ask if there are several $\, A(z)$ such that (\ref{mad})
 is verified. This is sketched in \ref{rota}.}. One can easily verify
that this series actually reduces (as it should!) to the successive 
iterates (\ref{iterR}) of $-4\, z/(1-z)^2$ for $\, a_1 \, = \, (-4)^n$.
In other words this one-parameter family of ``functions''
actually reduces to rational functions for an infinite number
of integer values  $\, a_1 \, = \, (-4)^n$. 

Furthermore, one can also verify a quite essential property
 we expect for a representation of the renormalization group
 namely that two  $\, R_{a_1}(z)$ for different values of $\, a_1$ commute, 
the result corresponding to the product of these two  $\, a_1$:
\begin{eqnarray}
\label{commute}
R_{a_1}\Bigl(R_{b_1} (z) \Bigr) \,\,  \, = \, \, \, \, 
R_{b_1}\Bigl(R_{a_1} (z) \Bigr) 
\,\,  \, = \, \, \, \, R_{a_1 \cdot b_1} (z).
\end{eqnarray}
The neutral element must necessarily correspond to $\, a_1 \, = \, 1$ 
which is actually the identity transformation $\, R_{1}(z) \, = \, \, z$.
We have an ``absorbing'' element  corresponding to $\, a_1 \, = \, 0$,
namely  $\, R_{0}(z) \, = \, \, 0$.
Performing the inverse  of $\, R_{a_1}(z)$ (with respect to the composition 
of functions)  amounts to changing 
$\, a_1$ into its inverse $\, 1/a_1$. 
Let us explore some ``reversibility'' property of our exact representation
of a renormalization group 
with the inverse of the rational transformations
(\ref{iterR}). The inverse of  $ R_{-4}(z)\, = \, \, -4\, z/(1-z)^2$
must correspond to $\, a_1 \, = \, -1/4$:
\begin{eqnarray}
R_{-1/4}(z)\, = \, \, - {\frac {1}{4}} \cdot z \, \,
-{\frac {1}{8}} \,{z}^{2} \, -{\frac {5}{64}}\,{z}^{3} \, 
-{\frac {7}{128}}\,{z}^{4}\, 
-{\frac {21}{512}}\,{z}^{5} \,\,  + \,\, \cdots 
\end{eqnarray}
However, a straight calculation of the inverse of 
 $ R_{-4}(z)\, = \, \, -4\, z/(1-z)^2$
gives a multivalued function, or if one prefers, two functions:
\begin{eqnarray}
\label{goodbranch}
&&S_{-1/4}^{(1)}(z)\, = \,\,  \, {\frac {z-2 \, +2\,\sqrt {1-z}}{z}}
 \, = \, \,\,  \,  \, 
 - {\frac {1}{4}} \cdot z \, \,
-{\frac {1}{8}} \,{z}^{2} \, + \,  \,  \cdots, \\ 
&&S_{-1/4}^{(2)}(z)\, = \, \, \,  {\frac {z-2 \, -2\,\sqrt {1-z}}{z}}
 \, = \, \,\,   \,  \, 
- {{4} \over {z}} \,  \, +2 \, + {{1} \over {4}} \,z \, 
+{{1} \over {8}}\,{z}^{2} \, + \, \cdots, \nonumber
\end{eqnarray}
which are the two roots of the simple quadratic
 relation ($R_{-4}(z') \, = \, \, z$):
\begin{eqnarray}
 z'^2 \, \, -  2\cdot  (1 \, -{{2} \over {z}}) \cdot z' \, \, + 1 
\,\,  = \, \,  \,\, 0, 
\end{eqnarray}
where it is clear that the product of 
these two functions is equal to $\, +1$.
The radius of convergence of $\, S_{-1/4}^{(1)}(z)$
 is $\, R \, = \, 1$.

Because of our choice to seek for functions analytical at $\, z\, = \, 0$
our renormalization group 
representation ``chooses'' the unique root that is 
analytical at $\, z\, = \, 0$, namely  $\,  S_{-1/4}^{(1)}(z)$.
For the next iterate of $ R_{-4}(z)\, = \, \, -4\, z/(1-z)^2$
in (\ref{iterR}) the inverse transformation 
corresponds to the roots of the 
polynomial equation of degree four ($R_{16}(z') \, = \, \, z$):
\begin{eqnarray}
z'^4\,\, +(4- {{16} \over {z}}) \cdot z'^3\, \,
 +(6+ {{32} \over {z}})  \cdot  z'^2 \,
 +(4- {{16} \over {z}}) \cdot z'\, \,  +1\,\, = \,\, \, 0, 
\end{eqnarray}
which yields four roots, one of which is analytical at
 $\, z \, = \, 0$ and corresponds to 
$\, a_1\, = \, \, 1/(-4)^2$ in our
 one-parameter family of (renormalization) 
transformations:
\begin{eqnarray}
S_{1/16}^{(1)}(z)\, = \,\,  \, 
 {{1} \over {16}} \,z \, +{\frac {3}{128}}\,{z}^{2} \, 
+{\frac {53}{4096}}\,{z}^{3} \,
 +{\frac {277}{32768}}\,{z}^{4} \, 
+{\frac {3181}{524288}}\,{z}^{5}\, + \, \, \cdots, \nonumber 
\end{eqnarray}
its (multiplicative) inverse  
$\, S_{1/16}^{(2)}(z)\, = \, \, 1/S_{1/16}^{(1)}(z)$:
\begin{eqnarray}
 S_{1/16}^{(2)}(z) \, = \, \, {{16} \over {z}} \, \, 
-6 \, -{\frac {17}{16}}\,z \, -{\frac {67}{128}}\,{z}^{2} \, 
-{\frac {1333}{4096}}\,{z}^{3} \, -{\frac {7445}{32768}}\,{z}^{4}
\, \, + \, \cdots \nonumber 
\end{eqnarray}
and two (formal) Puiseux series ($u \, = \, \,\pm \sqrt{z}$):
\begin{eqnarray}
S_{1/16}^{(3)}(z) \, = \,\, \,  \,
1 \, \, +u \,\,  +{{1} \over {2}} \,{u}^{2} \,\,
  +{{3} \over {8}} \,{u}^{3}\, \, 
+{{1} \over {4}} \,{u}^{4}\,  \, +{\frac {27}{128}}\,{u}^{5} \,\, 
 +{\frac {5}{32}}\,{u}^{6} \,+ \, \cdots \nonumber 
\end{eqnarray}
Many of these results are better understood when
 one keeps in mind that there is a special 
transformation $\, J: \, \, z \, \leftrightarrow \, 1/z$
 which is {\em also a R-solution of} 
(\ref{mad}) and verifies many compatibility 
relations with these transformations 
($Id$ denotes the identity 
transformation $\, R_{0}(z)$):
\begin{eqnarray}
&&R_{-4} \cdot J\,  \, = \, \, \,  \, R_{-4},
 \qquad \quad \quad  
 S^{(2)}_{-1/4} \cdot R_{-4} \,  \, = \, \, \,J, 
 \nonumber  \\
&& R_{-4} \cdot S_{-1/4}^{(1)}  \,  \, = \, \,  \, \, \,  
 S_{-1/4}^{(1)} \cdot R_{-4} \, \,  = \, \, \, Id,
 \qquad \nonumber  \\
&&S_{1/16}^{(1)}(z)\, \,  = \, \, \,  \, 
S_{-1/4}^{(1)} \cdot S_{-1/4}^{(1)}, \qquad \quad  
S_{1/16}^{(2)}(z)\, \,  = \, \, \, \,
 S_{-1/4}^{(1)} \cdot S_{-1/4}^{(2)},\nonumber  \\
&& J \cdot S_{-1/4}^{(1)}  \,  \, = \, \,   \,\,S_{-1/4}^{(2)},
\quad \quad   \qquad 
 J \cdot S_{-1/4}^{(2)}  \,  \, = \, \, \, \,S_{-1/4}^{(1)},
\qquad  \cdots \nonumber 
\end{eqnarray}
where the dot corresponds, here, to the composition of functions.
These symmetries of the linear differential operator $\, \Omega$
correspond to isogenies of the elliptic curve (\ref{ellip}).

It is clear that we have another one-parameter family 
 corresponding to $\, J \cdot R_{a_1}$ with an expansion of the form:
\begin{eqnarray}
\label{JR}
&&J \cdot R_{a_1} \, = \, \, \, \,
{{b_1} \over {z}} \, \, -{{2} \over {5}} \cdot (b_1\, -1)
\,  \,   -{{1} \over {15}} \cdot {{b_1^2 \, -1} \over {b_1}} \cdot z  
\nonumber \\
&&\quad 
 -{{2} \over {975}} \cdot 
{{(b_1 \, -1)\, (4\, b_1 \, +1)\, (4\,b_1 \, +3)  } \over {b_1^2}} \cdot z^2 
 \nonumber \\
&& \quad  -{{1} \over {248625}} \cdot 
{{(b_1 \, -1)\, (4\, b_1 \, +1)\, (1268\, b_1^2 \, +951\,b_1 \, +91)  }
 \over {b_1^3}} \cdot z^3 \nonumber \\
&&\quad
-{{2} \over {2071875}} \cdot 
{{(b_1 \, -1)\, (4\, b_1 \, +1)\, 
(3688\, b_1^3 \,+\, 2766\, b_1^2 \, +404\,b_1 \, +17)  }
 \over {b_1^4}} \cdot z^4
\nonumber \\
&&\quad \quad + \, \cdots 
\nonumber 
\end{eqnarray}
For $\, b_1\, =\, -1/4$,  
$\, b_1\, =\, (-1/4)^2$, $\, b_1\, =\, (-1/4)^3$, 
this family reduces to the (multiplicative) inverse
 of the successive rational functions
displayed in (\ref{iterR})
\begin{eqnarray}
&&-\, {{1} \over {4}} \cdot {{(1-z)^2} \over {z}} 
\quad \longrightarrow  \quad
{{1} \over {16}} \cdot {{(1+z)^4} \over {(1-z)^2 \cdot z}} 
\quad \longrightarrow  \quad \nonumber \\
&&\quad \longrightarrow  \quad
  -{{1} \over {64}} \cdot 
{{(1\, -6\,z\, +z^2)^4} \over {(1-z)^2 \cdot (1+z)^4 \cdot z}}
 \quad \longrightarrow  \quad
\cdots, \nonumber 
\end{eqnarray}
which can also be written as:
\begin{eqnarray}
&&-{{1} \over {4}} \cdot (z+{{1} \over {z}}) \, + \, {{1} \over {2}},
 \qquad \qquad  {{1} \over {16}} \cdot (z+{{1} \over {z}}) 
\, + \, {{3} \over {8}}\,
+\, {\frac {z}{ (1-z)^{2}}}, \nonumber \\
&& -{{1} \over {64}} \cdot (z+{{1} \over {z}})
 \,\, + \, {{13} \over {32}}\,\,
-{{z} \over {4}} \cdot {\frac {
17-60\,z+102\,{z}^{2}-60\,{z}^{3} +17\,{z}^{4} }
{ (1-z)^2 \, (1+ z)^4}}, 
\nonumber \\
&&{{1} \over {256}} \cdot (z+{{1} \over {z}}) \,\,\,
 + \, {{51} \over {128}}\,\,\,
+\,{{z} \over {16}} \cdot {\frac {
 17-60\,z+102\,{z}^{2}-60\,{z}^{3} +17\,{z}^{4}}{ (1-z)^2 \, (1+ z)^4}}
\nonumber \\
&& \quad \quad \quad
+16\,\,{\frac {z \cdot (1-z)^2 \, (1+ z)^{4}}{
 \, (z^2 \, -6\,z\, +1)^{4}}}, 
\nonumber \\
&&-{{1} \over {1024}} \cdot 
(z+{{1} \over {z}}) \, \, + \, {{205} \over {512}}\,\, 
-\,{{z} \over {164}} \cdot 
 {\frac {17 -60\,z+102\,z^2 -60\,z^3 +17\,z^4 }
{ (1-z)^2 \, (1+z)^{4}}}
\nonumber \\
&&\quad \quad \quad
-4\,{\frac {z \cdot (1-z)^2 \, (1+z)^4}{
 \, (z^2\, -6\,z \, +1)^{4}}}\, 
-64\,\,{\frac {z \cdot (1-z)^2 \, (1+z)^4
 \left( {z}^{2}-6\,z+1 \right)^{4}}
{ (1+20\,z-26\,{z}^{2} +20\,{z}^{3}+{z}^{4})^{4}}}, 
\nonumber \\
&&\qquad \cdots,  \nonumber \\
&&{{1} \over {(-4)^n}} \cdot (z+{{1} \over {z}}) \, \, 
+ \, {{2} \over {5\, 4^n}}\,(4^n -(-1)^{n} ) \, \, 
 \nonumber \\
&& \quad  \quad 
+{{z} \over {(-4)^{n-2}}} 
\cdot \,{\frac {17-60\,z+102\,{z}^{2}-60\,{z}^{3} +17\,z^4 }
{ (1-z)^{2} \, (1+ z)^{4}}}\nonumber \\
&&\quad  \quad 
+{{z} \over {(-4)^{n-6}}} \cdot 
\,{\frac { (1-z)^2 \, (1\, + z)^4}{
 \left( {z}^{2}-6\,z+1 \right)^{4}}}  \\
&& \quad  \quad 
+{{z} \over {(-4)^{n-8}}} \cdot
\,{\frac { (1-z)^{2} \, (1+z)^{4}
 \, ( {z}^{2}-6\,z+1)^{4}}
{ (1+20\,z -26\,{z}^{2} +20\,{z}^{3} +z^4)^4}}
\,\,\,\, + \,\,\, \cdots,  \nonumber
\end{eqnarray}
where we discover some ``additive structure''
 of these successive rational functions.

In fact, due to the specificity of this elliptic curve
 (occurrence of complex multiplication), we have
 another remarkable rational transformation 
solution of (\ref{mad}), preserving covariantly $\, \Omega$.
 Let us introduce the
rational transformation ($\, i$ denotes $\, \sqrt{-1}$):  
\begin{eqnarray}
\label{Tz}
T(z) \, = \, \, \, \,
 z \cdot \Bigl( {{z \, -(1\, +2\, i) }
 \over { 1 \, - \, (1\, +2\, i)\cdot z}} \Bigr)^4, 
\end{eqnarray}
we also have the remarkable covariance~\cite{Vidunas}:
\begin{eqnarray}
&&_2F_1\Bigl( [{{1} \over {2}}, {{1} \over {4}}],
 [{{5} \over {4}}];\,  z\Bigr) 
 \,  = \, \,\, \,
{{1 \, - \, z/(1+2\, i)  } \over {1 \, -(1+2\, i) \, z  }} 
 \cdot \, \, \, 
_2F_1\Bigl( [{{1} \over {2}}, {{1} \over {4}}], 
[{{5} \over {4}}]; \, T(z) \Bigr), 
\nonumber 
\end{eqnarray}
which can be rewritten in a simpler way on (\ref{defF}) (see (\ref{F})).

It is a straightforward matter to see that $\, T(z)$ actually belongs to
the $\, R_{a_1}(z)$ one-parameter family: 
\begin{eqnarray}
&&T(z) \, = \, \, R_{a_1}(z) \, = \, \, \, 
 -(7+24\, i) \cdot z \, \, + \, \, \cdots, \quad 
 \qquad  a_1 \, = \, \, -25 \cdot \rho, \quad \nonumber \\
&& \qquad \rho \, = \, \,(7+24\, i)/25, \qquad 
 \quad |\rho| \, = \, 1. \nonumber
\end{eqnarray}

As far as the reduction of (\ref{orderbyorder})  
to a rational function 
is concerned, 
it is straightforward to see that:
\begin{eqnarray}
\label{restrict}
&&(1-z)^2  \cdot (1+z)^4  \cdot R_{a_1}(z) 
\,\, = \, \, \, \,  \, 
 a_1 \cdot z \,\,  \, + \, \cdots \\
&&\,  \, 
-\, {{2} \over {175746796875}}  \cdot a_1 \cdot ( a_1\, -1)
\cdot ( a_1\, +4) \cdot ( a_1\, -16) \cdot
 P_8(a_1)  \cdot z^8\nonumber \\
&& \qquad  \, + \, \cdots  \nonumber \\
&&-\, {{1} \over {N(n)}}  \cdot a_1 \cdot ( a_1\, -1)
\cdot ( a_1\, +4) \cdot ( a_1\, -16) \cdot P_n(a_1) \cdot z^n
 \nonumber \\
&& \qquad \, 
+ \, \cdots \nonumber 
\end{eqnarray}
where $\, N(n)$ is a large integer growing with $\, n$, 
 and  $\, P_n$ is a polynomial with integer coefficients 
of degree $\, n\, -4$, 
or 
\begin{eqnarray}
\label{restrict2}
&& (1 \, - \, (1\, +2\, i)\cdot z)^4  \cdot R_{a_1}(z) 
\,\, = \, \, \, \, 
 a_1 \cdot z \,\, \,  + \, \cdots \, \nonumber \\
&&- \, {{4} \over {1243125}} \cdot a_1 \cdot (a_1\, -1)
 \cdot (a_1\, +7 \, +24 \, i) \cdot 
 (P_6(a_1)\, + i \, Q_6(a_1)) \cdot z^6 \nonumber \\
&& \qquad \, +  \, \cdots  \\
&&+ \, {{1} \over {N(n)}} \cdot a_1 \cdot 
(a_1\, -1) \cdot (a_1\, +7 \, +24 \, i)
 \cdot  (P_n(a_1)\, + i \, Q_n(a_1))
 \cdot z^n \nonumber \\
&& \qquad \, + \, \cdots \nonumber 
\end{eqnarray}
where  $\, P_n$ and  $\, Q_n$  are two
 polynomials with integer coefficients 
of degree respectively $\, n\, -3$ and $\, n-4$. 
 
Similar calculations can be performed for
  $\, T^{*}(z)$ defined by
\begin{eqnarray}
\label{Tbar}
T^{*}(z)\, = \, \,  z \cdot \Bigl( {{z \, -(1\, -2\, i) }
 \over {(1-2\, i) \, z \, -1 }} \Bigr)^4, 
\end{eqnarray}
for which we also have the  covariance:
\begin{eqnarray}
&&_2F_1\Bigl( [{{1} \over {2}}, {{1} \over {4}}],
 [{{5} \over {4}}];\,  z\Bigr) 
 \, \,  = \, \,\, \,\, 
{{1 \, - \, z/(1-2\, i)  } \over {1 \, -(1-2\, i) \, z  }} 
 \cdot \, \, \, 
_2F_1\Bigl( [{{1} \over {2}}, {{1} \over {4}}], 
[{{5} \over {4}}]; \, T^{*}(z) \Bigr). 
\nonumber 
\end{eqnarray}

It is a simple calculation  to check  that any iterate of $\, T(z)$,
(resp. $\, T^{*}(z)$) is actually a solution of (\ref{mad}) 
and corresponds to $\, R_{a_1}(z)$
for the infinite number of values 
$\, a_1 \, =\, \, \, (-7-24\, i)^N$ (resp. $(-7+24\, i)^N$).
Furthermore, one verifies, 
as it should (see (\ref{commute})), that the three
rational functions $\, R_{-4}(z)$, $\, T(z)$, 
and $\, T^{*}(z)$ commute. It is also
a straightforward calculation to see that
 the rational function built from
any composition  of $\, R_{-4}(z)$,
 $\, T(z)$ and $\, T^{*}(z)$ is actually a solution of (\ref{mad}).
We thus have a {\em triple infinity} of values of $\, a_1$, namely 
$\, a_1\, = \, (-4)^M \cdot (-7-24\, i)^N \cdot (-7+24\, i)^P$ 
 {\em for any integer $\, M$,  $\, N$ and $\, P$},
 for which $\, R_{a_1}(z)$ reduces to rational functions.
 We are in fact describing (some subset of)
 the isogenies of the elliptic curve (\ref{ellip}),
and identifying these isogenies with a discrete subset
 of the renormalization group. Conversely,
 a functional equation like (\ref{mad})
 can be seen as a way to extend the $\, n$-fold composition of a rational 
function $\, R(z)$ (namely $\, R(R( \cdots R(z) \cdots ))$)
 to $\, n$ {\em any complex number}.

\subsection{Revisiting the one-parameter family of solutions of the
 ``Rota-Baxter-like''  functional equation.}
\label{symsol2}

This extension can be revisited as follows. Keeping
 in mind the well-known example of the
parametrization of the standard map
 $\, z \, \rightarrow \,4\, z \cdot (1-z)$ with 
$\, z \, = \, \sin^2(\theta)$, yielding
 $\, \theta \, \rightarrow \, 2 \, \theta$,
let us seek for a (transcendental) 
parametrization $\, z \, = \, P(u)$
such that   
\begin{eqnarray}
R_{-4}\Bigl(P(u) \Bigr) \, = \, \, P(-4\, u) 
\qquad \hbox{or:}  \quad \, \,\, \,
R_{-4}\, = \, \,  P \cdot H_{-4} \cdot P^{-1},
\end{eqnarray}
where $\, H_{a_1}$ denotes the scaling
 transformation $\, z \, \rightarrow \, \, a_1 \cdot z$
(here  $\, H_{-4}: \,\, z \, \rightarrow \, -4 \cdot z $)
and $\, P^{-1}$ denotes the inverse transformation 
of $\, P$ (for the composition).
One can easily find such a (transcendental) 
parametrization order by order:
\begin{eqnarray}
\label{P}
&&P(z) \, = \, \, z \, - {{2} \over {5}} \,{z}^{2} \, 
+{\frac {7}{75}}\,{z}^{3} \, 
-{\frac {82}{4875}}\,{z}^{4}\, 
+{\frac {1078}{414375}}\,{z}^{5} \,-{\frac {452}{1243125}}\,{z}^{6}
\,  \nonumber \\
&&\qquad \qquad +{\frac {57311}{1212046875}}\,{z}^{7}\,\,
-{\frac {1023946}{175746796875}}\,{z}^{8}
\,\,  + \, \cdots  
\end{eqnarray}
and similarly for its inverse (for the composition) transformation:
\begin{eqnarray}
\label{Q}
&&Q(z) \, = \, \,\,P^{-1}(z) \, = \,\,  \,\,z \, 
+{{2} \over {5}} \,{z}^{2}\, 
+{\frac {17}{75}}\,{z}^{3}+{\frac {244}{1625}}\,{z}^{4}+
{\frac {45043}{414375}}\,{z}^{5} 
\nonumber \\
&&\quad \quad
+{\frac {2302}{27625}}\,z^6 +{\frac {128941}{1939275}}\,z^7\, 
+ \, {\frac {15365176}{281194875}}\, z^8
\,\, +\,\,  \cdots  
\end{eqnarray}
This approach is reminiscent of the conjugation introduced in Siegel's 
theorem~\cite{Siegel,Siegel2,Almost}. 
It is a straightforward matter to see (order by order) 
 that one actually has
\begin{eqnarray}
R_{a_1}\Bigl(P(u) \Bigr) \, = \, \,
 P(a_1\cdot u) \qquad \hbox{or:}  \qquad 
R_{a_1}\, = \, \,  P \cdot H_{a_1} \cdot P^{-1}.
\end{eqnarray}
The structure of the (one-parameter) renormalization group 
and the extension of the composition of $\, n$ times a rational 
function $\, R(z)$ (namely $R(R( \cdots R(z) \cdots ))$)
 to $\, n$ {\em any complex number},
becomes a straight consequence of this relation. 
Along this line one can define some 
``infinitesimal composition'' ($\epsilon \, \simeq \, 0$):
\begin{eqnarray}
\label{R1pluseps}
R_{1\, + \, \epsilon}(z)\, = \,  \,  \, \, 
 P \cdot H_{1\, + \, \epsilon} \cdot P^{-1}(z)
\, = \, \, \,  z \, + \, \epsilon \cdot F(z) \,\,  + \, \cdots, 
\end{eqnarray}
where one can find, order by order, 
the ``infinitesimal composition'' function $\, F(z)$:
\begin{eqnarray}
\label{infinitesimcompo}
&&F(z) \, = \, \,   \, \, 
z\,  \, - {{2} \over {5}}\,{z}^{2}\, \, 
 -{{2} \over {15}}\,{z}^{3} \, -{\frac {14}{195}}\,{z}^{4}
-{\frac {154}{3315}}\,{z}^{5}-{\frac {22
}{663}}\,{z}^{6}  \nonumber \\
&&\quad \quad \quad \quad
 -{\frac {418}{16575}}\,{z}^{7}-{\frac {9614}{480675}}\,{z}^{8}
-{\frac {2622}{160225}}\,{z}^{9}  \, 
\,  + \,  \cdots 
\end{eqnarray}
It is straightforward to see, from (\ref{commute}),
 that the function $\, F(z)$
satisfies the following functional equations involving 
 a rational function $\, R(z)$ (in the one-parameter
 family $\, R_{a_1}(z)$): 
\begin{eqnarray}
\label{condcompo}
&& {{dR(z) } \over {dz}} \cdot F(z) \,\, = \, \,\, F(R(z)), \qquad 
 {{dR^{(n)}(z) } \over {dz}} \cdot F(z) \,\, = \, \,\, F(R^{(n)}(z)), 
\nonumber \\
&& \quad \hbox{where:} \qquad \qquad 
 R^{(n)}(z) \, = \, \, \, R(R(\cdots R(z)) \cdots ). 
\end{eqnarray}
 $F(z)$ cannot be a rational
 or algebraic function. 
Let us consider the fixed points of $ R^{(n)}(z)$. 
Generically $ \, {{dR^{(n)}(z) } \over {dz}}$
 is not equal to $\, 0$ or $\, \infty$
at any of these  fixed points. Therefore  one must
 have $F(z) \, =\, 0$ or $F(z) \, = \,\infty$
for the infinite set of these fixed points:  $F(z)$ cannot be a rational
 or algebraic function, it is a transcendental
 function, and similarly for the parametrization
function $\, P(z)$.
In fact, let us introduce the function 
\begin{eqnarray}
\label{change}
&&G(z) \, = \, \, (1-z) \cdot F(z),  \\
&&G(z) \, = \,\, \, z\, \,-\,  {{7} \over {5}} \,{z}^{2} \, 
+{\frac {4}{15}}\,{z}^{3}+{\frac {4}{65}}\,{z}^{4}
+{\frac {28}{1105}}\,{z}^{5}+{\frac {44}{3315}}\,{z}^{6}
+{\frac {44}{5525}}\,{z}^{7} \nonumber \\
&& \quad \quad \quad 
+{\frac {836}{160225}}\,{z}^{8}\, +{\frac {1748}{480675}}\,{z}^{9} 
 \, + \, \cdots \, \, + \, g_n \cdot z^n \, + \, \cdots \, \,\nonumber 
\end{eqnarray}
One actually finds that the successive  $\, g_n$ 
satisfies the very simple 
(hypergeometric function) relation:
\begin{eqnarray}
{{ g_{n+1} } \over {g_{n}}} \, = \, \, \, \, \, 
{{ 4 \, n \, -\,9  } \over { 4 \, n \,+\,1}}. 
\end{eqnarray}
The function $\, G(z)$ is actually the hypergeometric function
solution of the homogeneous operator
\begin{eqnarray}
D_z^2 \, \, 
+\, {{1} \over {4}}\,{\frac {13\,z-3}{z \cdot (1-z) }} \cdot D_z
\, \, +\, \,
{{3} \over {4}}\,\,
{\frac {6\,{z}^{2}-3\,z+1}{ (1-z)^{2}\cdot {z}^{2}}},
\nonumber 
\end{eqnarray}
or of the inhomogeneous ODE:
\begin{eqnarray}
4\, z \cdot (1-z) \cdot {{d \, G(z)} \over {dz}} 
\, \, +\, (9 \, z\, -3) \cdot G(z) \, -z \cdot (1-z)^2
 \,\,  = \,\,\,  \, 0.
\nonumber 
\end{eqnarray}
One deduces the expression of $\, F(z)$ as a hypergeometric function
\begin{eqnarray}
\label{Fhyper}
F(z) \, = \, \,  \, z \cdot (1-z)^{1/2} \cdot \, \, 
_2F_1\Big([{{1} \over {4}},\, 
{{1} \over {2}}],[{{5} \over {4}}];\, z\Bigr)\, \,  = \, \,  \, 
{{\partial R_{a_1}} \over {\partial a_1}}|_{a_1 \, = \, 1}.
\end{eqnarray}

Finally we get the linear differential operator annihilating $\, F(z)$
\begin{eqnarray}
\label{OmegaF}
&& \Omega_F \, = \, \, \, D_z^2 \, \, 
+ \, {{1} \over {4}}\cdot {\frac {5\,z-3}{z \left( 1-z \right) }} \cdot D_z 
\, \, +\, {{1} \over {4}}\cdot
 {\frac {3-6\,z+5\,{z}^{2}}{ \left( 1-z \right) ^{2}{z}^{2}}} 
\nonumber \\
&& \qquad \quad \, = \,\,\, \, \, D_z \cdot \Bigl( D_z \, -{{1} \over {4}}
 \cdot {{ 3 \, -\, 5 \, z} \over {z \cdot (1-z)}}\Bigr),
\end{eqnarray}
which is, in fact,  nothing but  $\, \Omega^{*}$  
 the {\em adjoint of} linear
 differential operator $\, \Omega$ (see (\ref{Omega})).
One easily checks\footnote[3]{Using the command ``dchange''  with
 PDEtools  in Maple.} that the second order differential equation
 $\,\Omega_F(y(z)) \, = \, \, 0$  transforms under the change of
variable $\, z \, \rightarrow \, \, -4\, z/(1-z)^2$ into 
the second order differential equation
$\, \Omega_F^{(R)}(y(z)) \, = \, \, 0$
with $\, \Omega_F^{(R)}\, = \, \, C(z)^2 \cdot \omega_F^{(R)}$
where  the unitary (monic) operator 
 $\, \omega_F^{(R)}$  is the conjugate of 
$\, \Omega_F$: 
\begin{eqnarray}
&&\omega_F^{(R)}\, \, = \, \, \, \, \, 
D_z^2 \, - \, \, {{1} \over {4}}\cdot 
{\frac {11\,{z}^{2}+30\,z+3}{z \cdot (1- z)  \, (1+ z) }} \cdot D_z
\, \,  \\
&&\qquad  +\, {{1} \over {4}} \cdot
 {\frac {3\, +12\,z \,+50\,z^2\,+12\,z^3\,+3\,z^4 }
{ z^2 \cdot (1-z)^2 \, (1+ z)^2}}
\nonumber \\
&& \qquad \quad \, = \, \, \, \, \, 
\Bigl({{1} \over {C(z)}}  \Bigl)  \, \cdot  \, D_z   \cdot   
\Bigl(D_z \, -{{1} \over {4}}  \, \cdot  \,
  {\frac {3-5\,z}{z \cdot (1- z) }}  \Bigl)
 \,  \cdot \,  C(z)\nonumber \\
&& \qquad \quad \, = \, \, \, \, \,
 \Bigl({{1} \over {C(z)}}  \Bigl)  \, \, 
 \cdot  \, \Omega_F   \cdot \,  C(z)\, = \, \, \, \, \,
 \Bigl({{1} \over {C(z)}}  \Bigl) 
 \, \,  \cdot  \, \Omega^{*}   \cdot \,  C(z).
\nonumber
\end{eqnarray}
with $\, C(z) \, = \, \, 1/R'(z)$ and the 'dot'
 denotes the composition of operators.
Actually, the factors in the adjoint  $\, \Omega^{*}$ 
transform under the change of
variable $\, z \, \rightarrow \, \, -4\, z/(1-z)^2$  as
 follows\footnote[1]{Note that the result
 for $\, \omega_1^{*}$ is nothing but
transformation (\ref{omegaktrs}) on $\, \omega_k$
 for $\, k\, = \, \, -1$. Also note that the two transformations,
performing the change of
variable $\, z \, \rightarrow \, \, -4\, z/(1-z)^2$ and
 taking the adjoint, {\em do note commute}: 
 $\, (\omega_1^{*})^{(R)} \, \ne \, ((\omega_1)^{(R)})^{*}$.  }:
\begin{eqnarray}
&&D_z  \, \quad  \longrightarrow \, \,\quad C(z)\cdot  D_z, 
\qquad \quad
\omega_1^{*} \, = \, \, \, \quad  \longrightarrow \, \,\quad
 (\omega_1^{*})^{(R)}  \, = \, \, \,  \omega_1^{*} \cdot  C(z), 
\nonumber \\
&&\Omega^{*} \, \quad  \longrightarrow \, \,\quad 
 \Omega_F^{(R)}\, = \, \,\, \,
   \, C(z) \cdot \Omega^{*} \cdot C(z) \, 
\end{eqnarray}
which is precisely the transformation we need to match
 with (\ref{condcompo}) and see
the ODE $\, \Omega^{*}(F(z)) \, = \, \, 0$ compatible with the change of
variable $\, z \, \rightarrow \, \, -4\, z/(1-z)^2$:
\begin{eqnarray}
&&\Omega^{*}\, (F(z)) \, = \, \, 0 \, \quad  \longrightarrow \, \,\quad 
\Bigl(C(z) \cdot \Omega^{*} \cdot C(z) \Bigr)(F(R(z))) \,  \\
&&\qquad \, = \,\,  \,  \, 
\Bigl(C(z) \cdot \Omega^{*} \cdot C(z)\Bigr)(R'(z) \cdot F(z))
 \,\,  = \,\,  \, \, C(z) \cdot \Omega^{*}(F(z))
\,  \, = \,\,  \,  \, 0. \nonumber 
\end{eqnarray}
This is, in fact, a quite general result that will be seen to 
be valid in a more general (higher genus) framework 
(see (\ref{Fgenus5}), (\ref{adjgenus5}) below ). 

Not surprisingly one can deduce from (\ref{commute}) 
and the previous results, in particular
(\ref{Fhyper}), the following results for $\, R_{a_1}(z)$:
\begin{eqnarray}
\label{infinite }
&&\quad \quad -4 \cdot
 {{\partial R_{a_1}} \over {\partial a_1}}|_{a_1 \, = \, -4} 
\, \,  = \, \,  \,  \, 
F(R(z)), \nonumber \\
&&\quad \quad (-4)^n \cdot
 {{\partial R_{a_1}} \over {\partial a_1}}|_{a_1 \, = \, (-4)^n}
 \, \,  = \, \,  \,  \, 
F(R^{(n)}(z)), \nonumber 
\end{eqnarray}
where $R(z) \, = \, \, -4\, z/(1-z)^2$ and $\, R^{(n)}(z)$ 
denotes $\, R(R(\cdots R(R(z))))$. Of course we have similar relation 
for $\, T(z)$, $\, -4$ being replaced by $\, -7-24 \, i$. 
Therefore the partial derivative  $\, \partial R_{a_1}/\partial a_1$
 that can be expressed in terms of hypergeometric functions
 for for a {\em double infinity} of values of $\, a_1$, namely
 $\, a_1  \, = \, \,(-4)^M \times (-7-24)^N$.

\vskip .1cm 
One can, of course, check, order by order, that 
(\ref{condcompo}) is actually verified for any function in the 
one-parameter family $\, R_{a_1}(z)$:
\begin{eqnarray}
\label{condcompo2}
{{dR_{a_1}(z) } \over {dz}} \cdot F(z) 
\,\, = \, \,   \, \,\, F(R_{a_1}(z)).
\end{eqnarray}
which corresponds to an infinitesimal version of (\ref{commute}).

\vskip .3cm 

From (\ref{R1pluseps}) one simply deduces 
\begin{eqnarray}
\label{covP}
z \cdot {{dP(z)} \over {dz}} \,\, = \, \,   \, \,\, F(P(z)), 
\end{eqnarray}
that we can check order by order from (\ref{P}), the series expansion 
of $\, P(z)$, and from (\ref{infinitesimcompo})
the series expansion of $\, F(z)$,
but also 
\begin{eqnarray}
\label{QF}
{{d\, Q(z)} \over  {dz}} \cdot F(z)  \,\, = \, \,   \, \,\, Q(z) 
\end{eqnarray}
that we can, check order by order, from
 (\ref{Q}), the series expansion 
of $\, Q(z) \, = \, \, P^{-1}(z)$ and
 from (\ref{infinitesimcompo}). We now deduce that 
the log-derivative of the ``well-suited change of variable'' $\, Q(z)$
 is nothing but the (multiplicative) inverse of a hypergeometric
function $\, F(z)$:
\begin{eqnarray}
{{d \ln(Q(z))} \over  {dz}} \,\, = 
\, \,   \, \,\, {{1} \over {F(z)}}, \qquad 
Q(z) \, = \, \, \, 
\lambda \cdot exp\Bigl( \int^z {{dz} \over {F(z)}}\Bigr)
\end{eqnarray}
The  function $\, Q(z)$ is solution of 
the {\em non-linear} differential equation
\begin{eqnarray}
&& -4\, z^2 \cdot (1-z)^2 \cdot  
\Bigl(Q \cdot Q^{(1)}\cdot Q^{(3)} 
 \, \,+ (Q^{(1)})^2 \cdot Q^{(2)} \,\, -2\, Q  \cdot (Q^{(2)})^2\Bigr)
 \nonumber \\
&&\,\qquad 
\, + \, z \cdot (3- 5\,z)  \, (1- z) \cdot Q^{(1)} \cdot 
\Bigl(Q \cdot Q^{(2)}\, \, - \, (Q^{(1)})^2 \Bigr) \nonumber \\
&&\,\qquad  + \, (5\,{z}^{2}\, -6\,z\,+3) \cdot Q \cdot   (Q^{(1)})^2 
\,\, \,\,= \, \,\,\, \,\, 0,   
\end{eqnarray}
where the $\, Q^{(n)}$'s denote the $\, n$-th derivative of $\, Q(z)$. 
At first sight $\, Q(z)$ would be a {\em non-holonomic}
 function, however, remarkably, it is a  {\em holonomic}
function solution of an order-five operator which factorizes 
as follows:
\begin{eqnarray}
\label{OmegaQ}
&&\Omega_Q \, = \, \, \, 
\Bigl(D_z \, + \, {\frac {3-5\,z}{ (1-z) \cdot z }}\Bigr)  \cdot 
\Bigl(D_z \, + \,{{3} \over {4}} \cdot {\frac {3-5\,z}{ (1-z) \cdot z }}\Bigr)
  \cdot \\
&& \qquad \times 
\Bigl(D_z \, + \,{{2} \over {4}} \cdot {\frac {3-5\,z}{ (1-z) \cdot z  }}\Bigr) 
 \cdot 
\Bigl(D_z \, + \,{{1} \over {4}} \cdot {\frac {3-5\,z}{ (1-z) \cdot z  }}\Bigr) 
 \cdot D_z, 
\nonumber 
\end{eqnarray}
yielding the exact expression of $\, Q(z)$ in terms of hypergeometric functions:
\begin{eqnarray}
&&Q(z) \, = \, \, z \cdot \, 
\Bigl( _2F_1([ {{1} \over {2}}, \,  {{1} \over {4}} ], 
\, [{{5} \over {4}}];\, \,  z ) \Bigr)^4
 \, \, = \, \, \,  \\
&&\qquad \qquad  \, \, = \, \, \,
{{z } \over {1\, -z}} \cdot \, 
\Bigl( _2F_1([{{1} \over {4}} , \,{{3} \over {4}} ], 
\, [{{5} \over {4}}];\, \, 
 -\, {{z } \over {1\, -z}} ) \Bigr)^4. \nonumber  
\end{eqnarray}
that is the fourth power of (\ref{defF}),  the differential 
operator (\ref{OmegaQ})
being  the symmetric fourth power of $\, \Omega$.
From (\ref{defF}) we immediately get the covariance of $\, Q(z)$:
\begin{eqnarray}
\label{covQz}
Q\Bigl( -\, {{4 \, z} \over {(1-z)^2}} \Bigr)
   \, \,\,  = \, \,\,  \, \, \, -4 \cdot Q(z).
\end{eqnarray}
and, more generally, $\, Q\Bigl(R_{a_1}\Bigl) \, = \, \, a_1 \cdot Q(z)$. 
Since $\, Q(z)$ and $\, F(z)$ are expressed in term of the same 
hypergeometric function, the relation (\ref{QF}) must be an identity
on that hypergeometric function. This is actually the case. This 
hypergeometric function verifies the ingomogeneous 
equation:
\begin{eqnarray}
&&4 \cdot z \cdot {{d {\cal H}(z) } \over {dz}}\,  \,
 + \, {\cal H}(z) \, \, -(1-z)^{-1/2} 
 \, \, = \, \, \, 0, \\
&&\qquad \qquad  \hbox{where:} \qquad \qquad 
  {\cal H}(z) \,  \, = \, \, \,
 _2F_1\Bigl([ {{1} \over {2}}, \,  {{1} \over {4}} ], 
\, [{{5} \over {4}}];\, \,  z \Bigr). \nonumber 
\end{eqnarray}

Recalling $\, Q(P(z)) \, = \, z$, one has the following 
functional relation on $\, P(z)$
\begin{eqnarray}
\label{circular}
P(z) \cdot \, \,  _2F_1\Big([{{1} \over {4}},\, 
{{1} \over {2}}],[{{5} \over {4}}];\, P(z)\Bigr)^4
 \,  \,  \, = \,\, \,  \,\, z. 
\end{eqnarray}

Noting that $\, Q(z^4)^{1/4} \, = \, \, \, {\cal F}(z^4)$ 
 (see (\ref{defF})) can be expressed in term
of an incomplete elliptic integral of the first kind
of argument $\, \sqrt{-1}$
\begin{eqnarray}
 z \, \, \cdot \, _2F_1\Big([{{1} \over {4}},\, 
{{1} \over {2}}],[{{5} \over {4}}];\, z^4\Bigr)\,  \, 
= \, \, \,  \, EllipticF(z, \, \sqrt{-1}), 
\end{eqnarray}
one can find that (\ref{circular}) rewrites on
$\, P(z)$ as  
\begin{eqnarray}
EllipticF(P(z)^{1/4},\, \sqrt{-1})\,\, \,  = \,\, \,  \, \, z^{1/4}, 
\end{eqnarray}
from which we deduce that the function $\, P(z)$ is nothing but a 
{\em Jacobi elliptic
 function}\footnote[2]{Denoted $\, JacobiSN$ in maple:
 $P(z) = \, (JacobiSN(z^{1/4},I))^4$.}
\begin{eqnarray}
\label{CLOSED}
P(z)\,\, \,  = \,\,\, \,  \,  \Bigl( sn(z^{1/4},\, \sqrt{-1})\Bigr)^4.
\end{eqnarray}
In \ref{miscellanonlin} we display a set of
 ``Painlev\'e-like'' ODEs\footnote[1]{As a (non-holonomic) elliptic function
$\, P(z)$ provides elementary examples~\cite{Ince} of non-linear ODEs
 with the Painlev\'e property (like the Weierstrass-P function).}
 verified by  $\, P(z)$.
From the simple non-linear ODE on the Jacobi elliptic sinus,
namely $\, S"\, +2 \cdot S^3\, =\, 0$, and the exact expression of
$\, P(z)$ in term of Jacobi elliptic sinus,
  one can deduce other {\em non-linear} ODEs
verified by the {\em non-holonomic} function $\, P(z)$ 
($P^{(1)}\,=\,  dP(z)/dz$, $\, P^{(2)}\, = \, \, d^2P(z)/dz^2$): 
\begin{eqnarray}
\label{elliptisinusODE1}
&& z^{3/2}\cdot (P^{(1)})^2\, \,  \,  -\, (1-P)\cdot P^{3/2}
\, \, \,  \,  = \, \, \,\,\,  \, 0,   \\
\label{elliptisinusODE2}
&& P^{(2)}\,  \,  -{{3} \over {4}} \cdot {{ (P^{(1)})^2  } \over {P}} 
\,  \,  + {{3} \over {4}} \cdot   {{P^{(1)}} \over {z}} \,+ \, \, \,  {{1} \over {2}} \cdot 
{{P^{3/2}} \over {z^{3/2}}}  \,\, \,  \,  = \,\,\, \,  \,  0. 
\end{eqnarray}

\vskip .1cm

\subsection{Singularities of  the Jacobi elliptic function $\, P(z)$.}
\label{singul}

Most of the results of this section, and to some extent, of the next one,
are straight consequences of the exact closed expression of $\, P(z)$ in
term of an elliptic function. Following the pedagogical
approach of this paper we will rather follow a heuristic approach
not taking into account the exact result (\ref{CLOSED}), to display 
simple methods and ideas that can be used beyond exact results on a specific
example.
 
From a diff-Pad\'e analyzis of the series expansion of $\, P(z)$,
 we got the sixty (closest to $\, z \, = \, \, 0$) singularities.  
In particular we got that $\, P(z)$ has a radius of convergence 
$\, R \, \simeq \, \, 11.81704500807 \, \cdots \,  $ corresponding to the
following (closest to $\, z \, = \, \, 0$)
 singularity $\, z \, = \, \, z_s$ of $\, P(z)$:
\begin{eqnarray}
\label{radius}
&&z_s\, = \,  \, \, 
-11.817045008077115768316337283432582087420697 \, \cdots \,\,
\nonumber  \\
&&\quad \quad  =  \, \,\,\, \,
(-4)  \cdot \, \,  _2F_1\Big([{{1} \over {4}},\, 
{{1} \over {2}}],[{{5} \over {4}}];\, 1\Bigr)^4  \,  =  \, \,\,\, \,
 - \, {{ 1 } \over { 16}} \cdot  {{ \pi^6 } \over { \Gamma(3/4)^8}}.  
\end{eqnarray}
This singularity
 corresponds to a pole of order four: $\, P(z) \, \simeq \, (z-z_s)^{-4}$.
The function $\, P(z)$ has many other singularities:
\begin{eqnarray}
\label{miscellaneous}
&& 3^4 \cdot z_s, \quad (161\, \pm \, 240\, i)\cdot z_s, \quad
(-7\,\pm \, 24\, i)\cdot z_s, \quad (-119 \,\pm \,120\, i) \cdot z_s,
\  \cdots \nonumber  \\
 && 5^4 \cdot z_s, \quad  (41 \pm 840\, i)\cdot z_s, \quad
(-527 \pm 336\, i) \cdot z_s,  \quad  (-1519 \pm 720\, i)\cdot z_s,
 \   \cdots \nonumber  \\
 && 7^4 \cdot z_s, \  (1241 \pm 2520\, i)\cdot z_s, \
(-567 \pm 1944\, i)\cdot z_s,  \   (-3479 \pm 1320\, i) \cdot z_s,
 \   \cdots  \nonumber
\end{eqnarray}

In fact, introducing $\, x$ and $\, y$ 
the real and imaginary part of these singularities in $\, z_s$ units, 
one finds out that they correspond to the double infinity of points
\begin{eqnarray}
\label{xy}
&&x \, =    \,  \,
 (m_1^2 \, -2 \,m_1 \, m_2 \, - m_2^2) \cdot
 ( m_1^2 \, +2 \,m_1 \, m_2 \, - m_2^2) ,   \nonumber  \\
&& y \, = \, \, 
4 \cdot m_1 \, m_2 \cdot  (m_2 \,-\,  m_1 ) 
\cdot  (m_2 \,+\,  m_1 ),  
\end{eqnarray}
where $\, m_1$ and $\, m_2$ are two integers, 
and they all lie on the intersection
of an infinite number
of genus zero curves indexed by 
 the fourth power of an integer $\, M\,= m^4$
 ($m \, = \,  m_1$ or $m \, = \,  m_2$):
\begin{eqnarray}
&&2^{12}\,M^4  \,\,  -2^{11}\cdot x \cdot M^3 \, \, 
-\, 2^7 \cdot (17\,{y}^{2}+14\,{x}^{2}) \cdot M^2 \,  
\nonumber  \\
&&  \quad \quad \quad 
- \,2^5 \cdot x \cdot ( 8\,{x}^{2}\, +7\,y^2)\cdot  M \, \, \, 
+{y}^{4} \,\, \,  = \, \, \, \, 0.
\end{eqnarray}
The parametrization (\ref{xy}) describes not only the poles of $P(z)$
when $\, m_1 + \, m_2$ is odd, but also the zeros of $P(z)$ when
$\, m_1 + \, m_2$ is even.
This (infinite)  proliferation of singularities confirms
 the non-holonomic character of $\, P(z)$.

These results are simply inherited from (\ref{CLOSED}). 
The zeros and poles of the elliptic sinus 
$\, sn(z, \, i)$ 
correspond to two lattice of periods.  
Denoting $\, {\cal K}_1$ and $\,{\cal K}_2$  the two periods
of the elliptic curve, the location 
of the poles and zeros reads respectively: 
\begin{eqnarray}
&&P_{n_1,n_2} \, = \, \, 
2 \, n_1 \cdot {\cal K}_1 \, + \, (2 \, n_2 \, + \, 1) \cdot {\cal K}_2, 
\qquad \\
&&Z_{n_1,n_2} \, = \, \,
 2 \, n_1 \cdot {\cal K}_1 \, + \, 2 \, n_2 \cdot {\cal K}_2,    \\
&& {\cal K}_1 \, = \, \,  
  {{ \pi^{3/2} } \over { 2^{3/2}}} \cdot  {{ 1 } \over { \Gamma(3/4)^2}},
        \qquad {\cal K}_2 \, = \, \, 
 (1-\sqrt{-1}) \cdot K_1 
\end{eqnarray}
making crystal clear the fact that we have complex multiplication for
this elliptic curve. 
The formula (\ref{xy}) just amount to saying that the 
poles and zeros of $\, sn(z^{1/4}, \, i)$ are located at
$\, P_{n_1,n_2}^4$ and  $\, Z_{n_1,n_2}^4$:
\begin{eqnarray}
&& P_{n_1,n_2}^4\, = \, \, -{{z_s} \over {4}} \cdot
 \Bigl( (2 \, n_1 \, +  \,2 \, n_2 \, + \, 1) + \, i \cdot \,(2\, n_2 \, +
\, 1) \Bigr)^4
\nonumber \\
&&Z_{n_1,n_2}^4 \, = \, \,-{{z_s} \over {4}} \cdot
 \Bigl( (2 \, n_1 \, +  \,2 \, n_2 \, + \, 1) + \, i \cdot \,2\, n_2
\Bigr)^4 .
\end{eqnarray}
The correspondence with (\ref{xy}) is $m_1 =\, n_1 \,+ \, 2 n_2 \,+ 1$,
$m_2 =\, -n_1$ for the poles and
$m_1 =\,n_1 \,+ \, 2 n_2 $, $m_2= \, -n_1$ for the zeros.

\vskip .3cm 
{\bf Remark:} let us consider the
 $\, a_1 \, \rightarrow \, \infty$ limit
 of the one-parameter series $\, R_{a_1}$ (see
(\ref{anal}), (\ref{orderbyorder})) rewriting  $\, R_{a_1}(z)$
 as $\, \tilde{R}_{b_1}(u)$
\begin{eqnarray}
\label{limita1infty}
\tilde{R}_{b_1}(u) \, = \, \, R_{a_1}(z), \qquad \hbox{with:} \qquad
 b_1 \, = \, \, {{1} \over {a_1}}, \qquad 
u \, = \, \, {{z} \over {b_1}}.
\end{eqnarray}
In the $\, a_1 \, \rightarrow \, \infty$ limit, that is the
$\, b_1 \, \rightarrow \, 0$ limit, one easily verifies,
 order by order in $\, u$, that  $\, \tilde{R}_{b_1}(u) $
becomes {\em exactly} 
 the transcendental parametrization function (\ref{P}):
\begin{eqnarray}
\tilde{R}_{b_1}(u) \, \quad \rightarrow  \quad \, \,  \, P(u) 
\qquad \quad \hbox{when} \qquad \quad 
b_1 \quad \rightarrow  \quad 0. \nonumber 
\end{eqnarray}
For  $\, a_1 \, = \, \, (-4)^n \,$ ($n \, \rightarrow \, \, \infty$), 
one finds that
the radius of 
convergence\footnote[5]{It is to the absolute value of the
 inverse of the image of 
by the $\, n$-th iterate of $\, S_{-1/4}^{(1)}$ 
 of $\, -1$.} 
of the $\, R_{a_1}(z)$
 series becomes in the $\, n\, \rightarrow \, \infty$ limit 
$\, R_n \,  \simeq \, \, z_s/4^n$, in agreement 
with (\ref{limita1infty}).

\vskip .1cm 
\subsection{$P(z)$ and an infinite number of rational 
transformations: the sky is the limit. }
\label{sky}

Note that some non-linear ODEs
 associated with $\, P(z)$ and displayed in \ref{miscellanonlin}, namely
(\ref{ClosetoPainlV}) and (\ref{nonlinorderone}),
 and the functional equation (\ref{circular}),
 are invariant by the change of variable
$ \, (P(z), \,z)  \, \rightarrow \,( -4 \, P(z)/(1-P(z))^2, \, -4 \, z)$. 
In fact (\ref{ClosetoPainlV}),
 (\ref{circular}), and (\ref{nonlinorderone}) are invariant by
 $ \, (P(z), \,z)  \, \rightarrow \,( -4 \, P(z)/(1-P(z))^2, \, -4 \, z)$,
but also 
 $ \, (P(z), \,z)  \, \rightarrow \,( -(1-P(z))^2/4/P(z), \, -\, z/4)$,
and also by  $\, (P(z), \,z)  \, \, \rightarrow \, (1/P(z), \, z)$.

The function $\, P(z)$ satisfies the functional equation:
\begin{eqnarray}
\label{eqfuncP}
 P(-4 \cdot z) 
\, \,  = \, \, \,  \,  \, -\, {{ 4 \, P(z)} \over {1\, -P(z)^2}},  
\end{eqnarray}
but also 
\begin{eqnarray}
\label{eqfuncPT}
&&P((-7-24\, i) \cdot z) 
\, \,  = \, \, \, \, T(P(z)),  \\
&&P((-7+24\, i) \cdot z) 
\,  \, = \, \,\,   \, T^{*}(P(z)),   \nonumber 
\end{eqnarray}
and, more generally, as can be checked
 order by order on series expansions,
 \begin{eqnarray}
\label{eqfuncPa}
P(a_1 \cdot z) \, \,  = \, \, \,  \,  \, R_{a_1}\Bigl( P(z)\Bigr).
\end{eqnarray}
For example, considering the ``good'' branch  (\ref{goodbranch})
 for the inverse of
 $\, -4 \, z/(1-z)^2$, namely $\,S_{-1/4}^{(1)}(z)$,  we
can even check, order by order, on the series
 expansions of $\, P(z)$ and $\,S_{-1/4}^{(1)}(z)$
 the functional relation:
\begin{eqnarray}
\label{eqfuncS1over4}
S_{-1/4}^{(1)}(P(z)) \, \,  = \, \, \,  \, 
  P\Bigl( -\, {{z} \over {4}} \Bigr).  
\end{eqnarray}
{\em valid for $\,|P(z)| < \, 1$ 
since the radius of convergence of $\, S_{-1/4}^{(1)}(z)$
 is} $\, R \, = \, 1$.

Recalling the  functional equations (\ref{eqfuncPT}) 
 it is natural to say that if $\, P(z)$ is singular at 
$\, z\, = \,z_s$, then, for almost all the rational functions,
 in particular $\, T(z)$ (resp. $\, T^{*}(z)$)
the $\, T(P(z))$ is also singular $\, z\, = \,z_s$,
 and thus, from (\ref{eqfuncPT}), $\, P(z)$ 
is also singular at 
$\, z\, = \, (-7\,\pm \, 24\, i)\cdot z_s$.
It is thus extremely natural to see the emergence of the 
infinite number of singularities in (\ref{miscellaneous}) of the form 
$\, z\, = \, (N_1 \,+ \,  i \cdot N_2) \cdot z_s$, as a consequence of 
 (\ref{eqfuncPa}) together with a reduction of the one-parameter series
$\, R_{a_1}(z)$ to a rational function
 for an infinite number of selected values
of $\, a_1$, namely the $N_1 \,+ \,  i \cdot N_2$ in (\ref{miscellaneous}).
This is actually the case 
for all the values displayed in (\ref{miscellaneous}).
For instance, for $\, a_1\, = \, 3^4\, = \, 81$ we get
 the following simple rational function:
\begin{eqnarray}
\label{R81}
R_{81}(z) \, = \, \, \, 
 z \cdot \Bigl({\frac {{z}^{2}+6\,z-3}{3\,{z}^{2}-6\,z-1}}\Bigr)^4, 
\end{eqnarray}
for which it is straightforward to verify that this rational transformation
commutes with $\, T(z)$, $\, T^{*}(z)$,
 $\, -4 \, z/(1-z)^2$, and is a solution of 
the Rota-Baxter-like functional equation (\ref{mad}).
The case  $\, a_1\, = \, 5^4\, = \, 625$
 in (\ref{miscellaneous}), is even simpler, 
since it just requires to compose $\, T(z)$ and $\, T^{*}(z)$
\begin{eqnarray}
\label{R625}
&&R_{625}(z) \,\,  \, = \, \,  \,\, T(T^{*}(z)) 
\, \, \, = \, \, \, \, T^{*}(T(z)) \,  \, \\
&&\qquad \, = \, \,  \,
 z \cdot \Bigl( {\frac {{z}^{2}-2\,z+5}{5\,{z}^{2}-2\,z+1}}\Bigr)^4 
\cdot  \Bigl({\frac {1-12\,z-26\,{z}^{2}+52\,{z}^{3}+{z}^{4}}{
1+52\,z-26\,{z}^{2}-12\,{z}^{3} +{z}^{4}}}\Bigr)^4, \nonumber 
\end{eqnarray}
which, again verifies (\ref{mad}) and commutes with all
 the other rational functions, 
in particular (\ref{R81}). 
We also obtained the rational function corresponding to
 $\, a_1\, = \, 7^4\, = \,2401$, namely:
\begin{eqnarray}
\label{R2401}
&&R_{2401}(z) \,\,  \, = \, \,  \,\,
 z \cdot \Bigl({{N_{2401}(z)} \over {D_{2401}(z)}}\Bigr)^4,
 \quad \quad \quad \quad \hbox{with:}\\
\label{palindrom}
&&N_{2401}(z) \,\,  \, = \, \,  \,\,
 z^{12} \cdot D_{2401}\Bigl({{1} \over {z}} \Bigr), 
 \quad \quad \quad \quad 
\hbox{and:} \\
&&D_{2401}(z) \,\,  \, = \, \, 
1\, +196\,z-1302\,{z}^{2}+14756\,{z}^{3}-15673\,{z}^{4}\, 
-42168\,{z}^{5}\nonumber \\
&&\quad \quad
 +111916\,{z}^{6}-82264\,{z}^{7}+35231\,{z}^{8}-19852\,{z}^{9}\,
 \nonumber \\
&&\quad \quad
+2954\,{z}^{10}+308\,{z}^{11}-7\,{z}^{12}.
\end{eqnarray}

The polynomial $\, N_{2401}(z)$ satisfies
 many functional equations, like, for instance 
(with $R_{-4}(z)\, = \, \, -4\, z/(1-z^2)$):
\begin{eqnarray}
\label{magic2401}
4^{12} \cdot D_{2401}\Bigl({{1} \over {R_{-4}(z)}}\Bigr)
\,\, \,  \, = \,\,  \,  \,\,
D_{2401}(z) \cdot D_{2401}\Bigl({{1} \over {z}}\Bigr)
\end{eqnarray}
and also:
\begin{eqnarray}
\label{magic2401bis}
(1-z)^{49}\cdot D_{2401}\Bigl(R_{-4}(z) \Bigr)^2 \,  
 \, = \, \, D_{2401}(z)^4\,\, -z^{49} 
\cdot D_{2401}\Bigl({{1} \over {z}}\Bigr)^4. 
\end{eqnarray}

We also obtained the rational function corresponding to
 $\, a_1\, = \, 11^4\, = \,14641 $, namely:
\begin{eqnarray}
\label{R14641}
&&R_{14641}(z) \,\,  \, = \, \,  \,\,
 z \cdot \Bigl({{N_{14641}(z)} \over {D_{14641}(z)}}\Bigr)^4,
 \quad \quad \quad \quad \hbox{with:}\\
\label{palindrom2}
&&N_{14641}(z) \,\,  \, = \, \,  \,\,
 z^{30} \cdot D_{14641}\Bigl({{1} \over {z}} \Bigr), 
 \quad \quad \quad \quad \hbox{and:} \\
&&D_{14641}(z) \,\,  \, = \, \,
1\, +1210\,z-33033\,{z}^{2}+2923492\,{z}^{3}+5093605\,{z}^{4}
\nonumber \\
&&\quad \quad -385382514\,{z}^{5}
+3974726283\,{z}^{6}-14323974808\,{z}^{7}\nonumber \\
&&\quad \quad
 +57392757037\,{z}^{8}-291359180310\,{z}^{9}+948497199067\,{z}^{10}
\nonumber \\
&&\quad \quad 
-1642552094436\,{z}^{11}  +1084042069649\,{z}^{12}+1890240552750\,{z}^{13}
\nonumber \\
&&\quad \quad
-6610669151537\,{z}^{14}+9712525647792\,{z}^{15}-8608181312269\,{z}^{16}
\nonumber \\
&&\quad \quad 
+5384207244702\,{z}^{17}-3223489742187\,{z}^{18}+2175830922716\,{z}^{19}
\nonumber \\
&&\quad \quad
-1197743580033\,{z}^{20}+387221579866\,{z}^{21} -50897017743\,{z}^{22}
\nonumber \\
&&\quad \quad-
7864445336\,{z}^{23}+5391243935\,{z}^{24}-815789634\,{z}^{25}
\nonumber \\
&&\quad \quad
 +28366041\,{z}^{26}
 -5092956\,{z}^{27}+207691\,{z}^{28}+2794\,{z}^{29}-11\,{z}^{30}. 
\nonumber 
\end{eqnarray}
and, of course, one can verify that 
$\, R_{14641}(z)$ actually commutes with
 $\, R_{-4}$, $\, R_{81}$, $\, R_{625}$,
$\, R_{2401}(z)$, and is a solution 
of the Rota-Baxter-like functional equation
(\ref{mad}). Similarly to $\, R_{2401}(z)$
 (see (\ref{magic2401}), (\ref{magic2401bis})), 
we also have the functional equations:
\begin{eqnarray}
\label{magic14641}
4^{30} \cdot D_{14641}\Bigl({{1} \over {R_{-4}(z)}}\Bigr)
\,\, \,  \, = \,\,  \,  \,\,
D_{14641}(z) \cdot D_{14641}\Bigl({{1} \over {z}}\Bigr), 
\end{eqnarray}
and also: 
\begin{eqnarray}
\label{magic14641bis}
&&(1-z)^{(4\cdot 30 \, +1)}\cdot D_{14641}\Bigl(R_{-4}(z) \Bigr)^2 \,   \\
&& \qquad  \qquad \qquad  \, = \, \, 
D_{14641}(z)^4\,\, -z^{(4\cdot 30 \, +1)}
 \cdot D_{14641}\Bigl({{1} \over {z}}\Bigr)^4. 
\nonumber 
\end{eqnarray}

Next we obtained the rational function 
corresponding to
 $\, a_1\, = \, 13^4\, = \, 28561$, which verifies (\ref{mad}) namely:
\begin{eqnarray}
\label{R28561}
&&R_{28561}(z) \,\,  \, = \, \,  \,\,
 z \cdot \Bigl({{N_{28561}(z)} \over {D_{28561}(z)}}\Bigr)^4,
 \quad  \hbox{with:} \quad N_{28561}(z) \,\,  \, = \, \,  \,\,
 z^{42} \cdot D_{28561}\Bigl({{1} \over {z}} \Bigr),
 \nonumber \\
\label{palindrom3}
&&N_{28561}(z) \,\,  \, = \, \,  \,\,
 z^{42} \cdot D_{28561}\Bigl({{1} \over {z}} \Bigr), 
 \quad \quad \quad \quad \hbox{and:} \\
&&D_{28561}(z) \,\,  \, = \, \, (1-22\,z+235\,{z}^{2}
-228\,{z}^{3}+39\,{z}^{4}+26\,{z}^{5}+13\,{z}^{6})
 \cdot D^{(36)}_{28561}(z)
\nonumber \\
&&D^{(36)}_{28561}(z) \,\,  \, = \, \, 
1+2388\,z-61098\,{z}^{2}+19225300\,{z}^{3}+606593049\,{z}^{4}
\nonumber \\
&&\quad  -1543922656\,{z}^{5}+7856476560\,{z}^{6}-221753896032\,{z}^{7}+
1621753072244\,{z}^{8}\nonumber \\
&&\quad -4542779886736\,{z}^{9} +2731418674664\,{z}^{10}
+36717669656304\,{z}^{11}\nonumber \\
&&\quad -200879613202428\,{z}^{12}
+547249607666784\,{z}^{13}-934179604482832\,{z}^{14}\nonumber \\
&&\quad +1235038888776160\,{z}^{15}
-1788854212778642\,{z}^{16}
+3018407750933816\,{z}^{17}\nonumber \\
&&\quad -4349780716415868\,{z}^{18} +4419228090228152\,{z}^{19}
-2899766501472914\,{z}^{20}\nonumber \\
&&\quad
+931940880451552\,{z}^{21}+413258559018224\,{z}^{22} -857795672629664\,{z}^{23}
\nonumber \\
&&\quad 
+659989056851972\,{z}^{24}
-304241349909008\,{z}^{25}+87636987790824\,{z}^{26}\nonumber \\
&&\quad 
-14593362219920\,{z}^{27}+1073204980340\,{z}^{28}+45138167200\,{z}^{29}
\nonumber \\
&&\quad -23660433008\,{z}^{30}
+2028597792\,{z}^{31}-29540327\,{z}^{32}+3238420\,{z}^{33}
\nonumber \\
&&\quad -73386\,{z}^{34} -492\,{z}^{35}+{z}^{36}.
\end{eqnarray}
We get similar results, mutatis mutandis, than the ones previously obtained
 (commutation, functional equations like 
(\ref{magic14641}), (\ref{magic14641bis})...), namely:
\begin{eqnarray}
\label{magic28561}
4^{42} \cdot D_{28561}\Bigl({{1} \over {R_{-4}(z)}}\Bigr)
\,\, \,  \, = \,\,  \,  \,\,
D_{28561}(z) \cdot D_{28561}\Bigl({{1} \over {z}}\Bigr), \qquad  \cdots 
\end{eqnarray}

The ``palindromic'' nature of (\ref{R81})
(\ref{R625}), (\ref{R2401}), (\ref{R14641}) and (\ref{R28561})
 (see (\ref{palindrom}), (\ref{palindrom2})), (\ref{palindrom3})) corresponds to 
the fact that these rational transformations commute with $\, J$:
\begin{eqnarray}
{{1 } \over {R_{81}(z)}} \,\,  \, = \, \,  \,\,
 R_{81}\Bigl({{1 } \over {z}}\Bigr),
 \qquad {{1 } \over {R_{625}(z)}} \,\,  \, = \, \,  \,\,
 R_{625}\Bigl({{1 } \over {z}}\Bigr), \qquad \cdots 
\end{eqnarray}
In fact, more generally, we have $\, R_{N^4}(1/z) \, = \, \,  1/ R_{N^4}(z)$
 for  $\, N$ any odd integer ($N\, = \, \, 9, \, 21, \cdots, \, \cdots$)
and  $\, R_{N^4}(1/z) \, = \, \,  R_{N^4}(z)$ for  $\, N$ any even integer.

From (\ref{miscellaneous}) one can reasonnably conjecture that
the fourth power of {\em any integer} will provide a new example 
of $\, R_{a_1}(z)$ being a rational function. The simple non-trivial example 
corresponds to the already found rational function:
\begin{eqnarray}
R_{16}(z) \,\,  \, = \, \,  \,\, 
16\cdot {\frac {z \cdot (1\, - z) ^{2}}{ \left( z+1 \right)^{4}}}.
\end{eqnarray}
We already have explicit rational functions for all values of $\, a_1$ 
of the form $\, N^4$ for $\, N \, = \, 2, \, 3, \cdots, \, 16$ and of course,
 we can in principle, build explicit rational functions
 for all the $\, N$'s product
 of the previous integers. 
 Along this line it is worth noticing that the coefficients of the series
 $\, R_{a_1}(z)$ are all integers when $\, a_1$ is the 
 fourth power of any integer.

We are thus starting to build an infinite number of (elementary) 
{\em commuting} rational transformations, 
{\em any} composition of these
 (infinite number of) rational transformations
 giving rational transformations 
satisfying (\ref{mad}) and preserving the
linear  differential operator $\, \Omega$. This set 
of rational transformations is a pretty large set!
Actually this set 
of rational transformations
 corresponds to the isogenies of the underlying elliptic function. 

The proliferation of the singularities of $\, P(z)$ corresponds to
this (pretty large ...) set of rational transformations. Recalling 
 (\ref{eqfuncS1over4}), 
 the previous singularity argument is {\em not valid}\footnote[2]{
If this previous singularity argument were valid we would have had
 singularities as close as possible to $\, z \, = \, 0$
(namely $z_s/(-4)^n$), yielding a zero radius 
of convergence. Similarly combining 
$T^{*}(z)$ and the inverse of $\, T(z)$ we
 would have obtained an infinite number
 of singularities on the circle of radius $\, |z_s|$.} for the 
(well-suited) inverse transformations ($\, S_{-1/4}^{(1)}(z)$, ...) 
 of these rational transformations
because  (\ref{eqfuncS1over4}) requires $\,| P(z)| <\,  1$
(corresponding to the radius of convergence of $\, S_{-1/4}^{(1)}(z)$)
and the singularity $\, z \, = \, \, z_s$ corresponds precisely to ``hit'' 
the value $\, P(z)\, = \, \, 1$. 

\vskip .1cm 
\subsection{Other examples of selected
 Gauss hypergeometric ODE's}
\label{other}

 For heuristic reasons we have focused on 
$\, A(z)\, = \,\, (3 \, - \, 5 \, z)/z/(1-z)/4$, but of course, 
one can find many other examples and try 
to generalize these examples.

For instance, introducing 
\begin{eqnarray}
A(z) \, = \, \, \,{{1} \over {6}} \cdot {{d\ln((1-z)^3\, z^5)} \over {dz}}
 \, = \,  \, \,\,\,\,
{{1} \over {6}} \cdot {\frac {5-8\,z}{(1-z) \, z }}, \nonumber 
\end{eqnarray}
the rational transformation 
\begin{eqnarray}
R(z) \, = \, \,\, -27\cdot {\frac {z}{ (1-4\,z)^{3}}},
\end{eqnarray}
verifies the ``Rota-Baxter-like'' functional relation (\ref{mad}). 
This example corresponds to the following
 covariance~\cite{Vidunas} on a 
 Gauss hypergeometric integral (of the $\, c \, = \, 1+b$ type,
see below): 
\begin{eqnarray}
&&_2F_1\Bigl( [{{1} \over {2}}, {{1} \over {6}}],
 [{{7} \over {6}}];\,  z\Bigr)
 \, = \, \,\,\,  \, 
{{z^{-1/6}} \over {6}}  \cdot 
\int_0^{z}\, t^{-5/6} \, (1-t)^{-1/2} \cdot  dt 
 \,  \\
&& \qquad \quad\quad   \,  = \, \, 
(1\, -4\, z)^{-1/2} \cdot \, \, \, 
_2F_1\Bigl( [{{1} \over {2}}, {{1} \over {6}}], [{{7} \over {6}}]; 
\,  -27\,{\frac {z}{ (1-4\,z)^{3}}}\Bigr).
 \nonumber 
\end{eqnarray}
which is associated with the elliptic curve:
\begin{eqnarray}
y^6 \, - (1-t)^3\cdot  t^5 \,\,  = \, \, \, 0. 
\end{eqnarray}
Another example (of the $\, c \, = \, 1+a$ type, see below) is 
\begin{eqnarray}
\label{Atiers}
A(z) \, = \, \, \,
{{1} \over {3}} \cdot {{d\ln((1-z)^2\, z^2)} \over {dz}}
 \, = \,  \,\,\, \, 
 {{2} \over {3}} \cdot {\frac {1-2\,z}{ (1-z)\, z }}, 
\end{eqnarray}
where the rational transformation 
\begin{eqnarray}
\label{Rtiers}
R(z) \, = \, \, {\frac {z \cdot (z-2)^{3}}{ (1-2\,z)^{3}}}
\, = \, \,\, 
-8\,z-36\,{z}^{2}-126\,{z}^{3}-387\,{z}^{4}\,\, + \,\, \cdots  
\end{eqnarray}
verifies the ``Rota-Baxter-like'' functional relation (\ref{mad}). 
This example corresponds to the following
 covariance~\cite{Vidunas}  on a 
 Gauss hypergeometric integral: 
\begin{eqnarray}
&&_2F_1\Bigl( [{{1} \over {3}}, {{2} \over {3}}],
 [{{4} \over {3}}];\,  z\Bigr)
 \, = \, \, \, \,\,
{{z^{-1/3}} \over {3}}  \cdot \int_0^{z}\, t^{-2/3} 
\, (1-t)^{-2/3} \cdot  dt 
 \,  \\
&& \qquad \quad \quad \quad  \,  = \, \, \, 
{{1} \over {2}} \cdot {{ 2\, -z} \over {1 \, -2\, z}} \cdot \, \, \, 
_2F_1\Bigl( [{{1} \over {3}}, {{2} \over {3}}], [{{4} \over {3}}];
 \, {\frac {z \cdot (z-2)^{3}}{ (1-2\,z)^{3}}} \Bigr), 
\nonumber 
\end{eqnarray}
which is associated with the elliptic curve:
\begin{eqnarray}
y^3 \,\, - (1-t)^2\cdot  t^2 \,\,\, = \,\,\, \, 0. 
\end{eqnarray}
Note that, similarly to the main example of the paper, there
exist many rational 
transformations\footnote[4]{Note a (small) 
misprint in formula (64) page 174
 of Vidunas~\cite{Vidunas}.} satisfying (\ref{mad}) that cannot
 be reduced to iterates of (\ref{Rtiers}), 
for instance: 
\begin{eqnarray}
\label{missprint}
&&T(z)\, = \,\, \,\, -27\cdot \, {\frac {z \cdot (1-z)
  \left( {z}^{2}-z+1 \right) ^{3}}{
 \left( {z}^{3}+3\,{z}^{2}-6\,z+1 \right)^{3}}} 
\, = \, \,\,\, \,  -27\,z\,\,\,  -378\,{z}^{2} \\
&&\, \qquad \qquad 
-3888\,{z}^{3}-34074\,{z}^{4}-271620\,{z}^{5}\, 
-2032209\,{z}^{6}\,  \, + \,\,\,   \cdots \nonumber 
\end{eqnarray}
One verifies immediately that (\ref{missprint}) 
actually verifies (\ref{mad})
with (\ref{Atiers}). Not surprisingly, the two rational
 transformations (\ref{Rtiers}) and (\ref{missprint}) commute. 

Another simple example with rational symmetries corresponds to
$\, \Omega \, = \, \, \, $
$(D_z\, \, +A(z))\cdot D_z$ with 
\begin{eqnarray}
\label{Asaoud}
A(z) \,\, = \, \,\, \,
  -\, {{1} \over {2}} 
\cdot {\frac {3\,z-1}{z \left( 1-z \right) }}
\,\, = \,\, \,\, \,{{1} \over {2}} 
 \cdot {{d \ln(z\cdot (1-\, z)^2)} \over {dz}}. 
\end{eqnarray}
It has the simple (genus zero) hypergeometric solution\footnote[9]{Of the
 $c \, = \, 1 \, +b$ type (see below).}:
\begin{eqnarray}
{\cal F}(z)\,\, = \, \,\,\, \,  z^{1/2} \, \cdot \, 
_2F_1\Bigl([1,\, {{1} \over {2}}],[{{3} \over {2}}]; z\Bigr)
\,\, = \, \,\,\, \, arctanh(z^{1/2}).
\end{eqnarray}
The linear differential operator $\, \Omega$ is covariant 
under the change of variable $\, z \, \rightarrow \, 1/z$
and $\, z \, \rightarrow \, R(z)$
 where\footnote[1]{The change of variable (\ref{Rsaoud})
can be parametrized with hyperbolic tangents: $z \, \rightarrow \, z'$ 
 with $\, z\, = \, \, \tanh(u)^2$,
$\,\,  z'\, = \, \, \tanh(2\, u)^2$. Note that
 $z \, \rightarrow \, 4 \cdot z/(1-z)^2$ is parametrized by  
$\, z\, = \, \, \tan(u)^2$,$\,\,\, z'\, = \, \, \tan(2\, u)^2$
 but $z \, \rightarrow \, -4 \cdot z/(1-z)^2$ is not  parametrized by  
trigonometric functions. }:
\begin{eqnarray}
\label{Rsaoud}
R(z) \,\,  = \, \,\,\,  \, 
{{4 \, z } \over {(1\, +z)^2}}. 
\end{eqnarray}
One can easily check that (\ref{Asaoud}) and (\ref{Rsaoud}) 
satisfy the functional equation (\ref{mad}). One also verifies
that (\ref{Asaoud}) and $\, z \, \rightarrow \, 1/z$ or the iterates
of  (\ref{Rsaoud}) satisfy the functional equation (\ref{mad}).
The solution of the adjoint operator are $\,\, (1-z) \cdot  z^{1/2}\, $
and 
\begin{eqnarray}
\label{adjsaoud}
&&  F(z) \, \,\, = \, \, \, \, \,
z \cdot (1-z) \cdot \,
 _2F_1\Bigl([1, \, {{1} \over {2}}],[{{3} \over {2}}]; \, z \Bigr)
  \\
&& \qquad \, \,\, = \, \, \, \, \, 
z^{1/2}  \, \cdot \, (1-z) \cdot
arctanh(z^{1/2}) \, \,\, = \, \, \,\, \, z\, \, 
-\, {{2} \over {3}} \,{z}^{2}\, \, -{{2} \over {15}} \,{z}^{3}\, 
\, \nonumber \\
&& \qquad \qquad -{\frac {2}{35}}\,{z}^{4} \, -{\frac {2}{63}}\,{z}^{5}\, 
 -{\frac {2}{99}}\,{z}^{6}\, 
-{\frac {2}{143}}\,{z}^{7}\,\, + \, \,\cdots \nonumber
\end{eqnarray}
One verifies, again, that (\ref{adjsaoud}) and (\ref{Rsaoud})
commute, (\ref{adjsaoud}) corresponding to the ``infinitesimal
composition'' of (\ref{Rsaoud}) (see (\ref{R1pluseps})). 
\vskip .2cm 

A first natural generalization amounts to keeping the remarkable
factorization (\ref{Omega}) which will, in
 fact, reduce the covariance of a second order
operator to the covariance of a first
 order operator\footnote[5]{Thus avoiding 
the full complexity (and subtleties)
 of the covariance of ODE's by
  algebraic transformations like 
modular transformations (\ref{fundmodular}).}. 
Such a situation occurs for Gauss hypergeometric
 functions $\, _2F_1 \Bigl( [a, \, b], [1+a];\,  z\Bigr)$
solution of the $(a, \, b)$-symmetric linear differential operator
\begin{eqnarray}
\label{Gaussab}
 z\cdot (1-z)\cdot D_z^2\, \,\,  
 +(c-(a+b+1)\cdot z)\cdot D_z \,\,  -a \cdot b,
\end{eqnarray} 
 as soon 
as\footnote[2]{See for instance (\ref{gauss}) in \ref{appgauss}.}
 $\, c \, = \, 1+a$. For instance 
\begin{eqnarray}
\label{Fa}
{\cal F}(z)\, = \, \,\,\,  z^{a}\,  \cdot \, \, 
 _2F_1 \Bigl( [a, \, b],\, [1\,+a];\,  z\Bigr),  
\end{eqnarray}
is an integral of a simple algebraic function and is
 solution with the constant function
of the second order operator:
\begin{eqnarray}
\label{Omegagen}
&&\Omega \, \, = \,\,  \, 
\Bigl( D_ z \, +{{ (a-b-1)\, z \, +1-a} \over {z \cdot  (1-z)}} \Bigr)
 \cdot D_z 
\,  \, \,  \\
&& \qquad \qquad \, = \, \, \, 
 \Bigl( D_ z \, \, + {{d\ln((1-z)^{b} \cdot {z}^{1-a}) } \over {dz}}
\Bigr) \cdot D_z,  \nonumber 
\end{eqnarray}
yielding a new $\, A(z)$:
\begin{eqnarray}
\label{Agen}
A(z) \, = \, \, \, \, {\frac {(1\,-a) \, +(a-b-1) \, z}{ (1-z)\cdot  z}}
\,\,  = \, \, \, \, \,{{1-a} \over {z}} \,\,  - \, \, {{b} \over {1\, -z}}.
\end{eqnarray}
The adjoint of (\ref{Omegagen}) 
 has the simple solution  $\, z^{1-a} \cdot (1-z)^b$:
\begin{eqnarray}
F(z) \, = \, \, \,\, 
 z \cdot (1-z)^{b} \cdot \, _2F_1\Bigl([a, \, b],\,[1\,+a]; \, z \Bigr). 
\end{eqnarray}

Due to the $(a, \, b)$-symmetry of (\ref{Gaussab}) we
 have a similar result for $\, c\, = \, 1\, +b$. The function
$\,{\cal F}(z)\, = \,$
$ \,\, z^{b} \cdot  \,  _2F_1 \Bigl( [a, \, b],\, [1\,+b];\,  z\Bigr)\, $
is solution of (\ref{Omegagen}) where $\, a$ and $\, b$ have been permuted: 
\begin{eqnarray}
\label{Omegagen2}
\Bigl(  D_ z \,
 +{{(b-a-1)\, z \, +1-b} \over {z \cdot  (1-z)}}  \Bigr) \cdot D_z 
\end{eqnarray}
yielding another $\, A(z)$:
\begin{eqnarray}
\label{Agen2}
A(z) \, = \, \, \, \, {\frac {(1\,-b) \, +(b-a-1) \, z}{ (1-z)\cdot  z}},
\end{eqnarray}
The adjoint of (\ref{Omegagen2}) has the solution $\, (1-z)^a \cdot z^{1-b}\, $
 together with the hypergeometric function:
\begin{eqnarray}
F(z) \, = \, \, \,\, 
 z \cdot (1-z)^{a} \cdot \, \,  _2F_1\Bigl([a, b],[1+b]; \, z \Bigr). 
\end{eqnarray}
where one recovers the previous result (\ref{adjsaoud}).

We are seeking for (Gauss hypergeometric) second order differential 
equations\footnote[3]{More generally in our models of lattice statistical 
mechanics (or enumerative combinatorics etc.) we are seeking for
(high order) globally nilpotent~\cite{bo-bo-ha-ma-we-ze-09} 
 operators that, in fact, factor
into  globally nilpotent operators of smaller order, which, for 
Yang-Baxter integrable models with a canonical elliptic parametrization,
 must necessarily ``be associated with elliptic curves''.
 \ref{appchi2} provides some calculations showing that the integral
 for $\, \chi^{(2)}$, the two-particle 
contribution of the susceptibility of
 the Ising model~\cite{wu-mc-tr-ba-76,nickel-99,nickel-00} is clearly, and 
 straightforwardly, associated with an elliptic curve.}
 with an {\em infinite number} of
 (hopefully rational, if not algebraic)
symmetries: this is another way to say that we are not looking for 
generic Gauss hypergeometric differential 
equations, but Gauss hypergeometric differential 
equations {\em related to elliptic curves}, and thus having
 an infinite set of such isogenies. 
We are necessarily in the framework where the 
two parameters $\, a$ and $\, b$ of 
the Gauss hypergeometric are {\em rational numbers} in order to 
have {\em integral of algebraic functions} (yielding
 globally nilpotent~\cite{bo-bo-ha-ma-we-ze-09}
second order differential operators). Let us
 denote by $\, D$ the common denominator 
of the two rational numbers $\, a\, = \, N_a/D$ 
and $\, b\, = \, \, N_b/D$, the 
 function (\ref{Fa}) is associated to a period of the algebraic curve:
\begin{eqnarray}
 y^D \,\, = \, \,\, (1-t)^{N_b} \cdot t^{D-N_a}. 
\end{eqnarray}
We just need to restrict to triplets of integers 
$(N_a, \,N_b, \, D)$ such that 
the previous curve is an elliptic curve.

Let us give an example (of the $\, c \, = \, 1+b$ type)
 that {\em does not}
 correspond  to a genus one curve, with
\begin{eqnarray}
&&_2F_1\Bigl( [{{1} \over {3}}, {{1} \over {6}}],
 [{{7} \over {6}}];\,  z\Bigr)
 \, = \, \,  \,\,
{{1} \over {6}} \cdot z^{-1/6} \cdot
 \int_0^{z}\, t^{-5/6} \, (1-t)^{-1/3} \cdot dt, 
 \, \nonumber 
\end{eqnarray}
which corresponds to the {\em genus two} curve:
\begin{eqnarray}
\label{y6}
y^6 \,\, - (1-t)^2\cdot  t^5 \, \, \,= \,\,  \,\, 0. 
\end{eqnarray}
Again one introduces  $\, A(z)$:
\begin{eqnarray}
\label{Agenus2}
A(z) \, = \, \,\,\, 
{{1} \over {6}} \cdot {{d\ln((1-z)^2\, z^5)} \over {dz}}
 \, = \, \, \, \,
{{1} \over {6}} \cdot {\frac {5-7\,z}{z \cdot (1-z) }}, \nonumber 
\end{eqnarray}
and seek for $\, R(z)$ as series expansions analytical 
at $\, z \, = \, 0$.
One gets actually, order by order, a one-parameter family:
\begin{eqnarray}
\label{orderbyorder2}
&&R_{a_1}(z) \, = \,\,  \,\, a_1 \cdot z\, \, \, 
-{{2} \over {7}}\,a_1\, \cdot (a_1 -1) \cdot  {z}^{2}\,  \\
&&\qquad \quad 
+{\frac {1}{637}}\, a_1\, \cdot 
(a_1 -1)\cdot \left( 17\, a_1\, -87 \right) \cdot  {z}^{3}\nonumber \\
&&\qquad \quad  -{\frac {2}{84721}}\, a_1\, \cdot (a_1 -1)\cdot 
 \, (113\, a_1^{2}-856\,a_1+3438)  \cdot {z}^{4}\, \nonumber \\
&& \qquad \quad -{\frac {1}{38548055}}\,a_1\, \cdot (a_1 -1)\cdot 
 \, ( 3674\,a_1^{3}+121194\,a_1^{2} \nonumber \\
&& \qquad \quad \qquad \quad \quad 
-552261\, a_1\,  +2095059)\cdot   {z}^{5}
\,\,  \,  + \, \cdots \,  \nonumber \\
&& \qquad \quad 
 + \, {{1+\epsilon_n} \over {N(n)}}
 \cdot a_1  \cdot (a_1\, -1) \cdot P_n(a_1) \cdot   {z}^{n} 
\, \,\,   + \,\, \cdots\nonumber
\end{eqnarray}
where $\, \epsilon_n \, = \, 0$ for $\, n$ odd and 
 $\, \epsilon_n \, = \, 1$  for $\, n$ even, and $\, N(n)$ is a
 (large) integer depending on $\, n$, and $\, P_n(a_1)$ 
is a polynomial with integer coefficients of degree $\, n-2$.
One easily verifies, order by order,
 that one gets a one-parameter family of transformations
commuting for different values of the parameter:
\begin{eqnarray}
\label{comm}
R_{a_1}\Bigl(R_{b_1}(z)\Bigr) \, = \, \,\, \,\,
 R_{b_1}\Bigl(R_{a_1}(z)\Bigr) 
\, \, = \,\, \,\, \, R_{a_1\, b_1}(z). \nonumber 
\end{eqnarray}

As far as the ``algorithmic complexity'' of this series
(\ref{orderbyorder2})  is concerned
it is worth noticing that the degree growth~\cite{growth}
 of the series coefficients,
is actually {\em linear} and not exponential
 as we could expect~\cite{Functional} at first sight.
Even if this series is transcendental, it
 is not a ``wild'' series. 

Seeking for selected values of $\, a_1$ such that the previous 
series (\ref{orderbyorder2}) reduces to a rational function 
one can try to reproduce the 
simple calculations (\ref{restrict}), (\ref{restrict2}),
but unfortunately ``shooting in the dark'' because
 we have no hint of a well-suited denominator (if any!)
like the polynomials in the lhs of (\ref{restrict}), (\ref{restrict2}). 

It is also worth noticing that if we slightly change  $\, A(z)$ into:
\begin{eqnarray}
\label{AgenusN}
A(z) \, = \, \,\, 
{{1} \over {N}} \cdot {{d\ln((1-z)^2\, z^5)} \over {dz}}
 \, = \, \, \, 
{{1} \over {N}} \cdot {\frac {5-7\,z}{z \cdot (1-z) }}, 
\end{eqnarray}
the algebraic curve  (\ref{y6})
becomes $\, y^N \,\, - (1-t)^2\cdot  t^5 \, \, \,= \,\,  \,\, 0$
which has, for instance genus five for
 $\, N\, = \, \, 11$, but genus zero for $\, N\, = \, \, 7$.
For any of these cases of (\ref{AgenusN}) one
 can easily get, order by order,
 a one-parameter series 
$\, R_{a_1}$ totally similar to (\ref{orderbyorder2}) 
with, again, polynomials  $\, P_n(a_1)$ of degree $\, n-2$. 

The first coefficient $\, a_2$ is, in general:
\begin{eqnarray}
a_2 \, = \, \, 
- {{ 2} \over {2\, N\, -5 }} \cdot a_1 \cdot (a_1\, -1).  
\end{eqnarray}

For the genus zero case, $\, N=\, 7$:
\begin{eqnarray}
&&a_2 \, = \, \,
 - {{ 2} \over {9 }} \cdot a_1 \cdot (a_1\, -1), 
\, \, \quad \,
a_3 \, = \, \, 
- {{ 1} \over {1296 }}  \cdot a_1 
\cdot (a_1\, -1) \cdot ( 127\, - \, a_1 ), 
 \nonumber \\
&&a_4 \, = \, \, - {{ 1} \over {134136 }}  \cdot a_1 \cdot (a_1\, -1) 
\cdot (254\, a_1^2 \, +\,185\, a_1 \, +7499), \, \, 
\cdots  \nonumber \\
&&
a_n \, = \, \, - {{ 1} \over {N(n) }}  
\cdot a_1 \cdot (a_1\, -1) \cdot P_n(a_1). 
\nonumber 
\end{eqnarray}
which corresponds to the solution:
\begin{eqnarray}
\label{2sur7}
{{2} \over {7}} \cdot 
\int_0^z \, z^{-5/7} \cdot (1-z)^{-2/7} \cdot dt
 \,  \,= \, \,  \, \,\,
z^{2/7} \cdot \, \, 
 _2F_1\Bigl([{{ 2} \over {7}},{{ 2} \over {7}}],
 [{{ 9} \over {7}}]; z   \Bigr).
\end{eqnarray}
Using the parametrization of the {\em genus zero curve}
\begin{eqnarray}
y \, = \, \, \, -{\frac { (u+1)^2 \cdot  u^5}{ (u+1)^7 \, -\, u^7}},
 \qquad \quad
t \, = \, \, -{\frac {u^7}{ (u+1)^7 \, -\, u^7}}, 
 \nonumber
\end{eqnarray}
one can actually perform 
 the integration (\ref{2sur7}) of $\, dt/y$ and get an alternative 
form of the hypergeometric function (\ref{2sur7}):
\begin{eqnarray}
&&\int_0^z \, z^{-5/7} \cdot (1-z)^{-2/7} \cdot dt \,\,  \, = \, \,\, \, 
 \int_0^u \, \rho(u)  \cdot du  \,\,  = \, \, \, 
 \int_0^v \, {\frac {v}{1 \, - \, v^7}} \cdot dv,  \nonumber \\
&&  \hbox{where:}    \qquad \, \, \,  
z \, = \, \, -{\frac {u^7}{ (u+1)^7 \, -\, u^7}},
 \qquad  \quad \rho(u) \, = \, \, \,
 {\frac { (u+1)^4 \cdot u}{(u+1)^7 \, -{u}^7}}, \nonumber \\
&& \hbox{and:}  \qquad \qquad 
 v \, = \, \, {{u} \over {1+u}}, \qquad \,\quad 
 z \, = \, \, {\frac {{v}^{7}}{{v}^{7}-1}}.
 \nonumber 
\end{eqnarray}
Except transformations like $\, v \,\rightarrow  \, \omega \cdot v$ 
(with $\, \omega^7 \, = \, \, 1$)
which have no impact on $\, z$, it seems difficult to find rational 
symmetries in this genus zero case. 

For $\, N=\, 11$ (genus five) the first successive coefficients read:
\begin{eqnarray}
&&a_2 \, = \, \, 
- {{ 2} \over {17 }} \cdot a_1 \cdot (a_1\, -1), 
 \quad  \nonumber \\
&&
a_3 \, = \, \,
 - {{ 1} \over {8092 }}  \cdot a_1 \cdot (a_1\, -1)
 \cdot (143\, a_1 \, +367),
 \nonumber \\
&&a_4 \, = \, \, - {{ 1} \over {206346 }}  
\cdot a_1 \cdot (a_1\, -1) 
\cdot (1186\, a_1^2 \, +2473\, a_1 \, +5011), \, \,\, \,\,\,
\cdots  \nonumber \\
&&
a_n \, = \, \, 
- {{ 1} \over {N(n) }}  \cdot a_1 
\cdot (a_1\, -1) \cdot P_n(a_1). 
\nonumber 
\end{eqnarray}
The ``infinitesimal composition'' function $\, F(z)$
 (see (\ref{R1pluseps}), (\ref{infinitesimcompo}) 
and (\ref{condcompo})) 
reads:
\begin{eqnarray}
\label{Fgenus5}
&&F(z) \, = \, \, \,  
{{\partial R_{a_1}} \over {\partial a_1}}|_{a_1 \, = \, 1} 
  \, = \, \, \, 
 z\, -{\frac {2}{17}}{z}^{2}\, -{\frac {15}{238}}{z}^{3}\,
 -{\frac {5}{119}}{z}^{4}\,-{\frac {37}{1190}}{z}^{5} 
\nonumber \\
&&\quad \quad  -{\frac {888}{36295}}{z}^{6}-{\frac {2183}{108885}}{z}^{7}\,
 -{\frac {4366}{258213}}{z}^{8}\,-{\frac {58941}{4045337}}{z}^{9}\, \\
&& \quad \quad  -{\frac {1807524}{141586795}}{z}^{10}\, 
\, -{\frac {46543743}{4106017055}}{z}^{11}\, 
-{\frac {5305986702}{521464165985}}{z}^{12} 
\, +\,\,  \cdots \nonumber 
\end{eqnarray}
and, again we can actually check that
 this is actually the series expansion of
the hypergeometric function
\begin{eqnarray}
 z  \cdot (1-z)  \cdot     \, \,   
  _2F_1\Bigl([1, \, {{15} \over {11}}],[{{17} \over {11}}];  z  \Bigr), 
\end{eqnarray}
solution of $\, \Omega^{*}$ the adjoint 
of the $\, \Omega$ linear differential operator
corresponding to this (genus 5) $\, N=\, 11$ case:
\begin{eqnarray}
\label{adjgenus5}
\Omega^{*} \, = \,\, \,\,
 D_z \cdot 
 \Bigl(D_z \, \,
+{{1} \over {11}} \cdot {\frac {7\,z\, -5}{z \cdot (1-z) }} \Bigr). 
\end{eqnarray}

We have similar results for (\ref{Fa}), (\ref{Omegagen}), (\ref{Agen}). 
As far as these one-parameter family of transformations $\, R_{a_1}$,
are concerned, the only difference between
 the generic cases corresponding to {\em arbitrary
genus} and {\em genus one} cases like (\ref{Atiers}),
 is that, in the generic higher genus case,
only a {\em finite number} of values of the parameter $\, a_1$
can correspond to  rational  functions. Note that this higher genus result
generalizes to the arbitrary genus Gauss hypergeometric functions (\ref{Fa})
and associated operators (\ref{Omegagen}) and
 function (\ref{Agen}). In this general case
 one can also get order by order a one-parameter family
 of transformations $\, R_{a_1}$ satisfying a
commutation relation  (\ref{comm}). 

Note that $\, R(z) \, =\, \, 1/z$ is 
{\em actually a solution of} (\ref{mad})
{\em for this genus-two example} (\ref{Agenus2}).
Along this line of selected $\, R(z)$ solutions of (\ref{mad}) 
many interesting subcases of this general case
(\ref{Fa}), (\ref{Omegagen}), (\ref{Agen})
 are given in \ref{misc}.

\vskip .1cm 

In our previous genus-one examples,  
with this close identification between the {\em renormalization group
and the isogenies of elliptic curves}, we saw that, in order to obtain 
linear differential operators covariant by an infinite number 
of transformations (rational or algebraic), we must restrict 
our second order Gauss hypergeometric differential operator 
to Gauss hypergeometric {\em associated to elliptic curves}
 (see \ref{appgauss} and \ref{appchi2}). Beyond this framework we
 still have one-parameter families (see (\ref{comm}))
but we cannot expect an infinite number 
of rational (and probably algebraic) transformations
to be particular cases of such families of transcendental
 transformations. 

\vskip .1cm

\section{Conclusion}
\label{concl}
\vskip .1cm

We have shown that several selected Gauss hypergeometric linear differential 
operators associated to elliptic curves and factorised into order one
 differential operators actually 
present an {\em infinite number of rational symmetries} that actually identify 
with the {\em isogenies of the associated elliptic curves}
that are perfect illustrations of {\em exact representations of the
 renormalization group}. We actually displayed all these calculations, results
 and structures because they are perfect examples of exact renormalization 
transformations. 
For more realistic models (corresponding to Yang-Baxter models with
elliptic parametrizations), the previous calculations and structures become  
more involved and subtle, the previous rational transformations
 being replaced by algebraic transformations
 corresponding to {\em  modular curves}. 
 For instance, in our models of lattice statistical 
mechanics (or enumerative combinatorics, etc.), we are often getting
 globally nilpotent linear differential
operators~\cite{bo-bo-ha-ma-we-ze-09} of quite high
 orders~\cite{High,bo-gu-ha-je-ma-ni-ze-08,ze-bo-ha-ma-05b,Holonomy,higher3} 
 that, in fact, factor
into  globally nilpotent operators of smaller
 orders\footnote[5]{Experimentally~\cite{bo-gu-ha-je-ma-ni-ze-08} 
and as could be expected from Dwork's 
conjecture~\cite{bo-bo-ha-ma-we-ze-09}, one often finds 
for these small order factors hypergeometric second order operators
and sometimes selected Heun functions~\cite{maier-05} 
(or their symmetric products).}, which, for 
Yang-Baxter integrable models with a
 {\em canonical elliptic parametrization},
 must necessarily ``{\em be associated with elliptic curves}''.
 \ref{appchi2} provides some calculations showing that the integral
 for $\, \chi^{(2)}$, the two-particle contribution of the susceptibility of
 the Ising model~\cite{wu-mc-tr-ba-76,nickel-99,nickel-00} is clearly and 
 straightforwardly associated with an elliptic curve.

 We wanted to highlight   
 the importance of {\em explicit constructions} in answering
difficult or subtle questions. 

All the calculations displayed in this paper are elementary calculations
given explicitly for heuristic reasons.
The simple calculations (in particular with the introduction of
 a simple Rota-Baxter like functional equation) should be seen as some
undergraduate training to more realistic renormalization 
calculations that will require a serious knowledge of
 fundamental modular curves, {\em modular forms}, 
Hauptmoduls, Gauss-Manin or Picard-Fuchs
 structures~\cite{Painleve,Fuchs}
and, beyond, some  knowledge of mirror
 symmetries~\cite{Candelas,LianYau,Doran,Doran2,Kratten} 
of Calabi-Yau manifolds,
 these {\em mirror symmetries} generalizing\footnote[3]{See for
 instance equation (1.9) of~\cite{LianYau}. Do note that the periods of
 certain K3 families (and hence the original Calabi-Yau family)
 can be described by the squares of the periods of the elliptic
 curves~\cite{LianYau}. The mirror maps of some K3 surface families are 
always reciprocals of some McKay-Thompson series associated to the
Monstruous Moonshine list of Conway and Norton, the 
mirror maps of these examples
 being always automorphic functions
 for {\em genus zero}~\cite{Doran,Doran2}.} 
the  Hauptmodul structure for elliptic curves.

\ack
One of us (JMM) thanks  D. Bertrand,  A. Enge, M. Hindry,
D. Loeffler,  J. Nekov\'ar, J. Oesterl\'e and J. Watkins, 
for fruitful discussions on modular curves,
modular forms and modular functions. He also thanks 
D. Mouhanna for large discussions on the 
renormalization group and A.Ramani and N. Witte for detailed
 discussions on the Painlev\'e equations.

\appendix
\vskip .1cm

\section{Comment on the Rota-Baxter-like functional equation (\ref{mad})}
\label{rota}
\vskip .1cm
We saw, several times, that the Rota-Baxter-like
 functional equation (\ref{mad})
is such that for a given $\, A(z)$ one gets a one-parameter family 
of analytical functions $\, R(z)$ obtained order by order by
 series expansion (see (\ref{orderbyorder}), (\ref{orderbyorder2})).
Conversely for a given $\, R(z)$, for instance
 $\, R(z)\, = \, \,-4\,z/(1-z)^2$,
let us see if $\, R(z)$ can come from a unique $\, A(z)$.
Assume that there are two $\, A(z)$ satisfying (\ref{mad2}) with
the same $\, R(z)\, = \, \,-4\,z/(1-z)^2$. We will denote
 $\, \delta(z)$ the difference of these two $\, A(z)$, and 
 we will also introduce $\, \Delta(z) \, = z \cdot \delta(z)$. 
It is a straightforward calculation to see that $\, \Delta(z)$
 verifies
\begin{eqnarray}
\label{func}
\Delta(z) \, = \, \,\,\,\,\,
 {{ 1+z} \over {1-z}} \cdot \Delta\Bigl( {{-4\, z} \over {(1-z)^2}}  \Bigr), 
\end{eqnarray}
which has, beyond  $\, \Delta(z)\, = \, \, 0$, at least one 
solution analytical at $\, z\, = \, \, 0$ that we can get order by order:
\begin{eqnarray}
\Delta(z) \, = \, \,\,\,\,
 1 \, + {{2} \over {5}} \,z \, +{\frac {22}{75}}\,{z}^{2}
+{\frac {394}{1625}}\,{z}^{3} +{\frac {262634}{1243125}}\,{z}^{4}
\,  \, \, + \, \, \, \cdots \nonumber 
\end{eqnarray}
It is straightforward to show from (\ref{func}), from  similar
arguments we introduced for (\ref{infinitesimcompo}) on the functional
equations (\ref{condcompo})  that $\, \Delta(z)$
is a transcendental function.

\section{Miscellaneous non-linear ODE's on $\, P(z)$}
\label{miscellanonlin}

From (\ref{covP}) one can get 
\begin{eqnarray}
&&F'(P(z)) \, = \, \, \, \, 1 \, + \, \, z \cdot {{P^{(2)}} \over {P^{(1)}}}, 
 \nonumber \\
&&F"(P(z)) \, = \, \, \, \, 
{{P^{(2)}} \over {(P^{(1)})^2}} \,\,  
+\,  z \cdot {{P^{(3)}} \over {(P^{(1)})^2}}\, \, 
- \,  z \cdot {{(P^{(2)})^2} \over {(P^{(1)})^3}} , \qquad \nonumber 
\end{eqnarray}
and from (\ref{OmegaF}), the linear second order ODE on $\, F(z)$,
 one deduces the third order non-linear
 ODE\footnote[1]{Using differential algebra tools
one can verify that (\ref{elliptisinusODE2})
 implies (\ref{third}).} on the 
(at first sight {\em non-holonomic})
 function $\, P(z)$:
\begin{eqnarray}
\label{third}
&& z  \cdot (5\,{P}^{2}-6\,P+3) \cdot  (P^{(1)})^{4}\, 
 \, \, \, 
  -P  \cdot  (5\,P-3) \cdot   (P-1) \cdot  (P^{(1)})^{3} 
\, \nonumber \\
&& \quad \quad 
- \, z  \cdot (P-1)\cdot P \cdot
 (5\,P-3) \cdot P^{(2)} \cdot  (P^{(1)})^{2}  \, \nonumber \\
&&\quad \quad 
+4\,{P}^{2} \cdot (P-1)^{2}  \cdot (P^{(2)}\,  +z \cdot P^{(3)})
 \cdot  P^{(1)} \,   \\
&&\quad \quad 
-4\,z \cdot (P^{(2)})^2 \cdot  P^2 \cdot (P-1)^2 \, \, = \, \, \, 0,
\nonumber
\end{eqnarray}
where the $\, P^{(n)}$'s denote the $\, n$-th derivative of $\, P(z)$.  
This third order non-linear ODE has a rescaling symmetry 
$\, z \, \rightarrow \, \rho \cdot z$, for any $\, \rho$, 
and, also,  an interesting symmetry, namely
 an invariance by $\, z \, \rightarrow \, z^{\alpha}$, for
 any\footnote [1]{Beyond diffeomorphisms
 of the circle: the parameter $\, \alpha$ can 
be a complex number.} value of $\, \alpha$.

In a second step, using  differential algebra tools,
and, more specifically, the fact that 
$P(Q(z)) \, = \, \, Q(P(z)) \, = \, \, z$ 
together with the linear ODE for $\, Q(z)$, 
 one finds the simpler second order 
non-linear ODE for $\, P(z)$: 
\begin{eqnarray}
\label{ClosetoPainlV}
P^{(2)} \, 
-\,  {{1} \over {4}}
 \cdot {{5\,P-3} \over {(P-1) \cdot P}} \cdot (P^{(1)})^{2}\, \, 
 + \,  {{3} \over {4}} \cdot  {{1} \over {z}} \cdot P^{(1)} 
\, \, = \, \,\,  \, 0, 
\end{eqnarray}
or 
\begin{eqnarray}
P^{(2)} \,
 - \, \Bigl(  {{3} \over {4}} \cdot  {{1} \over { P}} \, 
+ \,{{1} \over {2}} \cdot  {{1} \over { P \, -1}}  \Bigr)
  \cdot (P^{(1)})^{2}\,
 + \,  {{3} \over {4}} \cdot  {{1} \over {z}} \cdot P^{(1)} 
\, \, = \, \,\,  \, 0. \nonumber 
\end{eqnarray}
Note that, more generally, the second order non-linear ODE 
\begin{eqnarray}
 \label{ClosetoPainlVbis}
P^{(2)} \,
 - \, \Bigl(  {{3} \over {4}} \cdot  {{1} \over { P}} \, 
+ \,{{1} \over {2}} \cdot  {{1} \over { P \, -1}}  \Bigr)
  \cdot (P^{(1)})^{2}\,
 + \,   {{\eta} \over {z}} \cdot P^{(1)} 
\, \, = \, \,\,  \, 0,
\end{eqnarray}
yields (\ref{third}) for any value of the constant $\, \eta$.
The change of variable 
 $\, z \, \rightarrow \, z^{\alpha}$, changes the parameter 
 $\, \eta$ into $ \, 1 \,  \, + \alpha \cdot (\eta\, -1)$.
In particular the involution $\, z \, \leftrightarrow \, 1/z$ 
changes $\, \eta\, = \, 3/4$ into $\, \eta\, = \, 5/4$.

This non-linear ODE, looking like {\em  Painlev\'e V}, is actually
 {\em invariant} by the change of variable 
$ \, P \, \rightarrow \, -4 \, P/(1-P)^2$. It is, also, invariant 
by any rescaling $ \, z \, \rightarrow \, \lambda \, z$, like the
particular degenerate\footnote[3]{Having the movable-poles solutions:
$\,(\alpha^{\beta} \, +  z^{\beta})^2/(\alpha^{\beta} \, -  z^{\beta})^2$.}
subcase of  {\em  Painlev\'e V}:
\begin{eqnarray}
\label{PainlV}
 y" \, \, 
- \, \Bigl( {{1} \over {2 \, y}}   \, \,
 + \, {{1} \over {y \, -\, 1}} \Bigr)
  \cdot y'^2 \, \,\,+ {{1} \over {z}}  \cdot y'  \,\, =  \,\, \, 0.
\end{eqnarray}

With (\ref{covP}) we recover the ``Gauss-Manin'' idea of 
Painlev\'e  functions being seen
 as deformations of 
elliptic functions:
\begin{eqnarray}
\label{deform}
z \cdot {{dP(z)} \over {dz}} \,\, = \, \,   \, \,\, 
 P(z) \cdot (1-P(z))^{1/2} \cdot \, \, 
_2F_1\Big([{{1} \over {4}},\, 
{{1} \over {2}}],[{{5} \over {4}}];\, P(z)\Bigr).
\end{eqnarray}
or:
\begin{eqnarray}
\label{deform2}
 -2 \, z \cdot {{d \, \, arctanh\Bigl( (1-P(z))^{1/2}\Bigr)} \over {dz}}
 \,\, = \, \,   \, \,\, 
 \, \, 
_2F_1\Big([{{1} \over {4}},\, 
{{1} \over {2}}],[{{5} \over {4}}];\, P(z)\Bigr). \nonumber
\end{eqnarray}
In fact, recalling $\, Q(P(z)) \, = \, z$,  one also has the relation
\begin{eqnarray}
\label{circularapp}
P(z) \cdot \, \,  _2F_1\Big([{{1} \over {4}},\, 
{{1} \over {2}}],[{{5} \over {4}}];\, P(z)\Bigr)^4
 \,  \,  \, = \,\, \,  \,\, z, 
\end{eqnarray}
yielding with (\ref{deform}) the simple {\em non-linear order-one}
 differential equation 
\begin{eqnarray}
\label{nonlinorderone}
 z^3 \cdot (P')^4\, \, \,  - (1-P)^2 \cdot P^3 \,\,  = \, \, \, 0,
\end{eqnarray}
already seen with (\ref{elliptisinusODE1}), 
and that we can write in a separate way:
\begin{eqnarray}
\label{nonlinorderoneform}
{{ dP} \over {(1-P)^{1/2} \cdot P^{3/4} } } 
\, \,  \, \, \, = \, \, \, \,  \,\,
 {{ dz} \over {z^{3/4}}}. 
\end{eqnarray}
Note that $\, P(z^{4 \cdot (1-\eta)  })$ 
is actually solution of (\ref{ClosetoPainlVbis}). 

Equation (\ref{nonlinorderone})  has (\ref{circularapp}) as a solution
wbut in general the Puiseux series solutions $P_A(z)$
 of the functional equation ($\mu$ is a constant):
\begin{eqnarray}
\label{Puiseux}
&&P_{\mu}(z)^{1/4} \cdot \, \,  _2F_1\Big([{{1} \over {4}},\, 
{{1} \over {2}}],[{{5} \over {4}}];\, P_A(z)\Bigr)
 \,  \,  \, = \,\, \,  \,\, \mu \, + \, \, z^{1/4}, 
\qquad \quad \hbox{or:} 
\nonumber \\
&&P_{\mu}(z) \, \, = \, \, \, \, P\Bigl( (\mu\, + \, z^{1/4})^4  \Bigr), 
 \nonumber \\
&& P_{\mu}(z) \, \, = \, \, \, \, 
P(\mu^4) \, + \, \,\, 4 \cdot \mu^3 \cdot P'(\mu^4) \cdot z^{1/4} \,\, + \, \,
\cdots 
\end{eqnarray}
It is a straightforward exercise of differential algebra to see that
the order-one non-linear differential equation (\ref{nonlinorderone})
implies (\ref{ClosetoPainlV}). In particular not only (\ref{circularapp}) 
is solution of (\ref{ClosetoPainlV}) but also all the Puiseux series
solutions (\ref{Puiseux}) of (\ref{nonlinorderone}).
More generally the solutions of the functional equation:
\begin{eqnarray}
\label{Puiseuxscaled}
P_{\mu,\lambda}(z)^{1/4} \cdot \, \,  _2F_1\Big([{{1} \over {4}},\, 
{{1} \over {2}}],[{{5} \over {4}}];\, P_{\mu,\lambda}(z)\Bigr)
 \,  \,  \, = \,\, \,  \,\, \mu \, + \, \, \lambda \cdot z^{1/4}.
\end{eqnarray}
verify (\ref{ClosetoPainlV}). 
This corresponds to the fact that 
\begin{eqnarray}
\label{nonlinorderoneL}
 z^3 \cdot (P')^4\, \, \,  - \lambda^4 \cdot (1-P)^2 \cdot P^3 
\,\,  = \, \, \, 0,
\end{eqnarray}
yields (\ref{third}) which is scaling 
symmetric ($z \rightarrow \, \rho \cdot z$)
when (\ref{nonlinorderone}) is not. 
More generally 
\begin{eqnarray}
\label{nonlinorderoneLbis}
 z^{4\, \eta} \cdot (P')^4\, \, \,  - \lambda^4 \cdot (1-P)^2 \cdot P^3 
\,\,  = \, \, \, 0,
\end{eqnarray}
 yields (\ref{third}) for any value of
 the parameters $\, \eta$ and $\, \lambda$.
Finally, one also has that the solution of the functional equation
\begin{eqnarray}
\label{thirdsol}
P_{\eta}(z)^{1/4} \cdot \, \,  _2F_1\Big([{{1} \over {4}},\, 
{{1} \over {2}}],[{{5} \over {4}}];\, P_{\eta}(z)\Bigr)
 \,  \,  \, = \,\, \,  \,\, \mu \, + \, \, \lambda \cdot z^{1-\eta},
\end{eqnarray}
are solutions of (\ref{third}), but also of (\ref{ClosetoPainlVbis}) 
and even of (\ref{nonlinorderoneLbis}).

Equation (\ref{ClosetoPainlVbis}) with $\, \eta \, = \, 1/2$ 
(instead of  $\, \eta \, = \, 3/2$ in (\ref{ClosetoPainlV})), 
has a solution, analytical at $\, z \, = \, 0$ :
\begin{eqnarray}
\label{newseries}
&&1\, +x\, + {{1} \over {2}}\,  x^2 \, + {{7} \over {40}}\, x^3 \, 
+ \,{{1} \over {20}}\,x^4 \, 
+\, {{121} \over {9600}}\, x^5 \, 
+ \, {{7} \over {2400}}\, x^6\, \nonumber \\
&&\qquad +\, {{211} \over {332800}}\,x^7
+ \, {{41} \over {312000}}\, x^8 \, \, + \, \, \cdots 
\end{eqnarray}
This series has a singularity at $\,-1/4 \cdot z_s^2$ where
 $\, z_s$ is given by (\ref{radius}).
The radius of convergence of (\ref{newseries}) corresponds to this
 singularity, namely  $\,R \, = \, 1/4 \cdot z_s^2$.
This singularity result
can  be understood
from the fact , at $\, \eta \, = \, 1/2$,  $\, P(z^2)$ is 
actually solution of (\ref{ClosetoPainlVbis}).

In fact, we have the following solutions 
of (\ref{ClosetoPainlVbis}) for various selected values of $\, \eta$.
For $\, \eta \, = \, 0$, $\, P(z^4)$ is solution of (\ref{ClosetoPainlVbis}).
For $\, \eta \, = \, 2/3$, $\, P(z^{4/3})$
 is solution of (\ref{ClosetoPainlVbis}), and, more generally,
$\, P(z^{4 \cdot (1-\eta)})$ is solution of
 (\ref{ClosetoPainlVbis}).

\section{Gauss hypergeometric ODE's related to elliptic curves}
\label{appgauss}
\vskip .1cm
It is not necessary to recall the close connection between 
Gauss hypergeometric functions and elliptic curves,
 or even modular curves~\cite{Stiller,Zudilin} and 
Hauptmoduls. This is very clear on the Goursat-type relation
\begin{eqnarray}
\label{Haupt}
&&_2F_1 \Bigl([2\, a, {{2\, a+1} \over {3}}], 
[{{4\, a+2} \over {3}}];  x \Bigr)\, \, \,   = \, \,\, 
  \\
&& \quad \, \,\, \,\,  (1\, -x\,+\,  x^2)^{-a}\cdot \, \, 
 _2F_1 \Bigl([{{a} \over {3}}, {{a+1} \over {3}}], [{{4\, a+5} \over {6}}];
 {{ 27} \over {4}} \cdot {{ (x-1)^2 \cdot x^2} 
\over {(1\, -x\,+\, x^2)^3 }}  \Bigr),
 \nonumber 
\end{eqnarray}
which generalizes the simpler quadratic Gauss relation:
\begin{eqnarray}
_2F_1 \Bigl([a,b], [{{a+b+1} \over {2}}];  x \Bigr)
 \, = \, \,\, \, \,   \, 
 _2F_1 \Bigl([{{a} \over {2}}, {{b} \over {2}}], 
[{{a+b+1} \over {2}}];  4\, x\, (1-x) \Bigr).
 \nonumber 
\end{eqnarray}
On (\ref{Haupt}) 
one recognizes (the inverse of) the {\em Klein modular 
invariant}\footnote[1]{Taking for $\, x$ the elliptic lambda function.}
for the pull-back of the 
hypergeometric function on the rhs. 

Many values of $[[a,\, b], [c]]$ are known to correspond to  
 elliptic curves like $[[1/2,\, 1/2], [1]]$ (complete elliptic integrals
of the first and second kind) 
or  modular forms: 
$[[1/12,\, 5/12], [1]]$,  $[[2/3,\, 2/3], [1]]$,   $[[2/3,\, 2/3], [3/2]]$
 and they can even be simply related:   
\begin{eqnarray}
&&\, \Bigl( {{z+27} \over {27}} \Bigr)^{1/3} \,\cdot
 \, \,  _2F_ 1([2/3,\, 2/3], [\,1];\,-1/27\,z) 
\, = \, \, \nonumber \\
&& \quad \qquad \qquad \, 
\mu(z)\,
 \cdot \, 
_2F_1\Bigl([1/12,\, 5/12], [\,1];\,1728\,{\frac {z}{ \left( z+27 \right) 
 \left( z+3 \right)^{3}}}\Bigr), \nonumber \\
&&\hbox{where:} \qquad \qquad \mu(z) \, = \, \,  \, 
\Bigl(  {{(z+27)  \, (z+3)^{3}} \over {729}} \Bigr)^{-1/12}.
\nonumber\end{eqnarray}

Once we have a hypergeometric function corresponding to 
an elliptic curve for some values of $(a, \, b, \, c)$,
one can find other values of $(a, \, b, \, c)$ 
also corresponding to elliptic curves:
\begin{eqnarray}
_2F_1 \Bigl([a,b], [c]; x \Bigr)
 \,\,  \longrightarrow  \, \,\, \,\, 
 x^{1-c} \cdot \, \,   _2F_1 \Bigl([1+a-c, 1+b-c], [2-c]; x  \Bigr).
 \nonumber 
\end{eqnarray}

In order to provide simple examples of linear differential ODE's
we will restrict ourselves (just for heuristic reasons)
to Gauss hypergeometric second order differential equations. 

Let us recall the {\em Euler integral representation}
 of the Gauss hypergeometric functions:
\begin{eqnarray}
\label{gauss}
&&_2F_1\Bigl([a, \, b],\,  [c];\,  z \Bigr) \,  \\
&& \quad  \, = \, \,\, 
{{ \Gamma(c)} \over {\Gamma(b) \,\Gamma(c-b)  }} \cdot \int_0^1 \, 
{{dw} \over {w}} \, \, w^b \cdot (1-w)^{c-1-b}  \cdot (1-\, z\, w)^{-a}
\nonumber \\
&& \quad  \, = \, \,\,
{{ \Gamma(c)} \over {\Gamma(a) \,\Gamma(c-a)  }} \cdot \int_0^1 \, 
{{dw} \over {w}} \, \, w^a \cdot (1-w)^{c-1-a} 
 \cdot (1-\, z\, w)^{-b}
\nonumber \\
&& \quad  \, = \, \,\,
{{ \Gamma(c)} \over {\Gamma(a) \,\Gamma(c-a)  }} \cdot z^{-a} 
\int_0^z \, {{du} \over {u}} \, \,
 u^a \cdot (1-{{u} \over {z}})^{c-1-a} 
 \cdot (1-\, u)^{-b}. \nonumber 
\end{eqnarray}
\vskip .4cm 
On the last line of (\ref{gauss}), the selected
 role of $\,\,  c \, = \, 1+\, a\, $ is quite clear. 

Recall that the corresponding second order differential operator 
 is {\em invariant under the permutation of $\, a$ and $\, b$}
which is not obvious\footnote[2]{For instance for
 $\,2F_1([1/4, \, 1/2],\,  [5/4];\,  z)$ 
it changes an Euler integral  with 
${{ \Gamma(5/4)} \over {\Gamma(1/4) \,\Gamma(1)  }}
 \, = \, \,\, {{1} \over {4}}$
into an Euler integral  with
${{ \Gamma(5/4)} \over {\Gamma(3/4) \,\Gamma(1/2)  }} \, = \, \,\,
 {{1} \over {4}} \cdot {{ (2\, \pi)^{1/2}} \over {\Gamma(3/4)^2}}$.} 
on the Euler integral representations 
of the hypergeometric functions (this amounts to 
permuting $\, 0$ and $\, \infty$). The permutation of $\, a$ and $\, b$
 is always floating around in this paper.

When the three parameters  $\, a$, $\, b$ and $\, c$ 
of the Gauss hypergeometric 
functions are rational numbers we have
 integrals of {\em algebraic functions}
and, therefore, we 
know~\cite{bo-bo-ha-ma-we-ze-09,Andre,Andre2,Andre3,Andre4} 
that the corresponding second order 
differential operator is necessarily
 {\em globally
 nilpotent}~\cite{bo-bo-ha-ma-we-ze-09,Andre,Andre2,Andre3,Andre4}.
Let us restrict to  $\, a$, $\, b$ and 
$\, c$ being {\em rational numbers}
$\, a\, = \, N_a/D$, $\, b\, = \, N_b/D$ and $\, c\, = \, N_c/D$
where $\, D$ is the common denominator of these three rational numbers.
The Gauss hypergeometric functions are naturally associated to the 
{\em pencil of algebraic curves}:
\begin{eqnarray}
y^D  \, = \, \, \,\,  \, (1-u)^{N_b} \cdot u^{D-N_a} 
\cdot \Bigl( 1-{{u} \over {z}} \Bigr)^{-N_c\, + D \, +N_a}.
 \end{eqnarray}
Recalling the main example of the paper, one associates with 
$\, _2F_1\Bigl([1/4, \, 1/2],\,  [5/4];\,  z \Bigr)$
\begin{eqnarray}
&&_2F_1\Bigl([1/2, \, 1/4],\,  [5/4];\,  z \Bigr)
 \,\,  = \, \,\,  \nonumber \\
&& \, = \, \,\,
 {{ \Gamma(5/4)} \over {\Gamma(1/2) \,\Gamma(3/4)  }} \cdot
 \int_0^1 \, 
{{dw} \over {w}} \, \cdot  w^{1/2} \cdot (1-w)^{-1/4} 
 \cdot (1-\, z\, w)^{-1/4}
\nonumber \\
&& \, = \, \,\,   {{ \Gamma(5/4)} \over {\Gamma(1/2) \,\Gamma(3/4)  }}
 \cdot  z^{-1/2} \cdot \int_0^z \, 
  u^{-1/2} \cdot (1\, -{{u} \over {z}})^{-1/4}  \cdot (1-\, u)^{-1/4} \cdot du 
\nonumber
\end{eqnarray}
 the $\, z$-pencil of   elliptic curves\footnote[1]{The algebraic 
curves (\ref{pencil}) are genus
 one curves for any value of $\, z$, except $\, z\, = 1$
 where the curve becomes the union of two rational 
curves $\, (u^2-u+y^2)\, (u^2-u-y^2)\, = \, 0$.} 
\begin{eqnarray}
\label{pencil}
{y}^{4} \,\,  -\, u^{2} \cdot (1-u) \cdot (1-{\frac {u}{z}})
\,  \, = \,\,  \, 0,
\end{eqnarray}
where we associated (see (\ref{ellip})) to 
$\,\, _2F_1\Bigl([1/2, \, 1/4],\,  [5/4];\,  z \Bigr)\,$ the elliptic curve
\begin{eqnarray}
 {y}^{4} \,\,  -\, u^{3} \cdot (1-u)^2 \,  \, = \, \, \,\, 0.
\end{eqnarray}

\subsection{Miscellaneous examples}
\label{misc}
\vskip .1cm

In the more general (\ref{Fa}), (\ref{Omegagen}), (\ref{Agen}),
(resp. (\ref{Omegagen2}), (\ref{Agen2}))  framework,
one can find many interesting subcases. 
\vskip .1cm 
$\bullet$ The previous $\, R(z) \, = \, \, 1/z$
involution is solution of the functional relation (\ref{mad})
when $\,\,  a\, = \, 2\, b\,\,$ if $\,\, c\, = \, 1 \, +b$,  or
 $\, \, b\, = \, 2\, a\,\,$ if $\,\, c\, = \, 1 \, +a$.
 
$\bullet$ The involution $\, R(z) \, = \, \, 1 \, -z$
is solution of the functional relation (\ref{mad})
when $\,  a\,+\, b\, = \,\, 1$ if $\,\, c\, = \, 1 \, +b$, or
 $\,\, c\, = \,\, 1 \, +a$.
 
$\bullet$ The infinite order transformation:
\begin{eqnarray}
R(z) \, = \, \,\,
 t \cdot {\frac {z}{ 1\, + \, (t-1)\cdot  z }}, 
\quad \quad \quad 
R^{(n)}(z) \, = \, \,\,
 t^n \cdot {\frac {z}{ 1\, + \, (t^n-1)\cdot  z }},
\nonumber 
\end{eqnarray}
is solution of the functional relation (\ref{mad})
when $\,  a\, = \, 1\, + \, b$ if $\, \, c\, = \, 1 \, +b$,  or
$\,  b\, = \, 1\, + \, a$
 $\, c\, = \, 1 \, +a$.

$\bullet$ The scaling transformation $\, R(z) \, = \, \, t \cdot z$
 is solution of the functional relation (\ref{mad})
when $\,\,  a \, = \, \, 0$ and $\,\,  c\, = \, 1 \, +b$ 
(resp. $\,\,  b \, = \, \, 0$ and $\,\,  c\, = \, 1 \, +a$).

$\bullet$ We also have a quite degenerate situation
 for $\,\,  b \, = \, \, 1$ or $\,\,  a\, = 1$ when $\, c\, = \, 2$ 
with the infinite order transformation
\begin{eqnarray}
R(z) \, = \, \, \, 1 \, \, - t \cdot (1-z), \quad \quad \quad 
R^{(n)}(z) \, = \, \,\, 1 \, \, - t^n \cdot (1-z), \nonumber 
\end{eqnarray}
solution of (\ref{mad}).

$\bullet$ The two order-three transformations
\begin{eqnarray}
R(z) \, = \, \,\, {\frac {z\, -1}{z}},
 \qquad \qquad 
R(R(z)) \, = \, \, {{1} \over {1-z}},  \nonumber 
\end{eqnarray}
are solutions of the functional relation (\ref{mad})
for $\, a =\,  2/3,\, \,\,   b = \, 1/3, \,\,  \, c\, = \, 4/3$,  or 
 $\,\, a =\,\,  1/3,\, \, b = \,\, 1/3, \,\, c\, = \,\, 4/3$.

\section{Ising model susceptibility :  $\, \tilde{\chi}^{(2)}$
and elliptic curves}
\label{appchi2}
\vskip .1cm

The two-particle contribution of the susceptibility of
 the Ising model~\cite{wu-mc-tr-ba-76,nickel-99,nickel-00} 
is given by a double integral. 
This double integral on two angles $\, \tilde{\chi}^{(2)}$ reduces to a simple
integral\footnote[2]{The prefactors in front of the integrals 
 are not relevant for our discussion here.} (because
 the two angles are opposite):
\begin{eqnarray}
\tilde{\chi}^{(2)} \, =  \, \, \, 
\int_0^{\pi} \, d\theta \cdot y^2 \cdot {{1 \, +x^2} \over {1 \, -x^2}}
\cdot \Bigl( {{x \cdot \sin(\theta)} \over {1\, -x^2}} \Bigr)^2,  
\nonumber 
\end{eqnarray}
where 
\begin{eqnarray}
&&x \, = \, \,\,  A \, - \, B, \qquad \quad
A \, = \, \,  {{1} \over {2\, w}} \, - \, \cos(\theta), 
\qquad \quad
B^2 \, = \, A^2\, -1, \nonumber \\
&&y^2  \, = \, \,{{1} \over { A^2\, -1}}.\nonumber 
\end{eqnarray}
Denoting $\, C\, = \, \, \cos(\theta)$ we can rewrite the
 integral $\, \chi^{(2)}$ as :
\begin{eqnarray}
\label{C0}
\tilde{\chi}^{(2)}  \, =  \, \, \, 
\int_0^{1} \, {{dC} \over {(1-C^2)^{1/2}}} \cdot x^2 \cdot y^2 \cdot
 {{1 \, +x^2} \over {(1 \, -x^2)^3}},   
\end{eqnarray}
that we want to see as:
\begin{eqnarray}
\label{C1}
\int_0^{1} \, {{dC} \over {z}} \, = \, \,\,
 \int_0^{w} \, {{dq} \over {Z}}. 
\end{eqnarray}
The variable $\, z$  reads: 
\begin{eqnarray}
{{1} \over {z}} \,\,  
-\,  {{1} \over {(1-C^2)^{1/2}}} \cdot x^2 \cdot y^2 \cdot
 {{1 \, +x^2} \over {(1 \, -x^2)^3}} 
\, = \, \, \,\,  0
\end{eqnarray}
which after simplifications gives
\begin{eqnarray}
A^2\, (C^2-1)\cdot z^2 \,  +(A^2-1)^5 \, = \,\, \, 0,
\end{eqnarray}
that is
\begin{eqnarray}
 ({{1} \over {2\, w}}\, -C)^{2} \cdot (C^{2} \, -1) \cdot  z^2
\,\,\, + \, (({{1} \over {2\, w}}\, -C)^{2} \, -1)^5 
\, = \,\,\,  \, \, 0.
\nonumber 
\end{eqnarray}
In terms of the variable $ \, q \, = \, \, w \cdot C$
one can rewrite (see (\ref{C1})) the integral (\ref{C0}) 
as an incomplete integral:
\begin{eqnarray}
 && 256 \cdot  (1-2\,q) ^{2} \, (q^2 -w^2)  
 \cdot Z^{2}{w}^{4} \nonumber \\
&& \qquad \qquad + \, (2\,q-1+2\,w)^{5}
 \, (2\,q-1-2\,w)^{5}
\, = \, \, \,\,\,  0. \nonumber 
\end{eqnarray}
This $\, w$-pencil of algebraic curves is actually a {\em $\, w$-pencil of
 genus one curves},  
seen as algebraic curves in $\, Z$ and $\, q$.


\section*{References}

\begin{thebibliography}{10}

\bibitem{Migdal}
http://www.physics.fsu.edu/courses/Spring05/phy6938-02/decimation.pdf

\bibitem{Fisher} M.E. Fisher, 
{\em Renormalization group theory: Its basis
 and formulation in statistical physics}, 
Reviews of Modern Physics, {\bf 70}, No. 2,  (1998) pp. 653--681

\bibitem{DSFisher} D. S. Fisher, {\em Random fields,
 random anisotropies, nonlinear
  $\, \sigma$ models, and dimensional reduction},
 Phys. Rev {\bf B 31}, 7233-7251  (1985) 

\bibitem{Wetter} J. Berges, N. Tetradis and C. Wetterich, 
	{\em Non-perturbative renormalization flow 
          in quantum field theory and statistical physics},
        Phys. Rep.  {\bf 363} (2002) 223 and arXiv:hep-ph/0005122

\bibitem{Delam} B. Delamotte, D. Mouhanna and  M. Tissier,
	{\em Nonperturbative renormalization-group 
          approach to frustrated magnets}, 
        Phys. Rev.  {\bf B 69} (2004) 134413

\bibitem{broglie}
    J-M. Maillard and S. Boukraa, 
	{\em Modular invariance in lattice statistical mechanics},
	 Annales de l'Institut Louis de Broglie, Volume 26,
	num\'ero sp\'ecial, pp. 287--328 (2001)

\bibitem{bo-ha-ma-ze-07b}
S. Boukraa, S. Hassani, J.-M. Maillard and N. Zenine,
{\em Singularities of $n$-fold integrals of the
 Ising class and the theory of elliptic curves},
J. Phys. A: Math. Theor. \textbf{40} (2007) 11713--11748 
and arXiv.org/pdf/0706.3367

\bibitem{Automorphisms} J-M. Maillard,
	{\em Automorphisms of algebraic varieties and Yang-Baxter equations},
        Journ. Math. Phys. {\bf 27}, (1986), pp. 2776--2781

\bibitem{Baxterization} S.~Boukraa and  J-M. Maillard,
	 {\em Let's Baxterise},  J. Stat. Phys. {\bf 102}, (2001), 641-700,
         and : arXiv: hep-th/0003212

\bibitem{Canada} B. C. Berndt and H. H. Chan, 
{\em Ramanujan and the Modular $j$-Invariant},
 Canad. Math. Bull. {\bf 42} (1999) 427--440

\bibitem{BeMaVi92}
	M.P. Bellon, J-M. Maillard and C-M. Viallet,
	{\em Quasi-Integrability of the sixteen vertex model}, 
	 Phys.Lett. {\bf B 281}, (1992), pp. 315--319 

\bibitem{bo-ha-ma-ze-07}
S. Boukraa, S. Hassani, J.-M. Maillard and N. Zenine,
 {\em Landau singularities and singularities of
 holonomic integrals of the Ising class},
 J. Phys. A: Math. Theor. {\bf 40} (2007) 2583--2614
 and  arXiv:math-ph/0701016 v2

\bibitem{ze-bo-ha-ma-04}
N. Zenine, S. Boukraa, S. Hassani and  J.-M. Maillard,
{\em The Fuchsian differential equation of the square 
 Ising model $\, \chi^{(3)}$ susceptibility},
J. Phys. A: Math. Gen. {\bf 37} (2004) 9651--9668 and  
arXiv:math-ph/0407060

\bibitem{ze-bo-ha-ma-05}
N. Zenine, S. Boukraa, S. Hassani and J.-M. Maillard,
{\em Square lattice Ising model susceptibility: series expansion method, 
and differential equation for $\chi^{(3)}$},
J. Phys. A: Math. Gen. {\bf 38} (2005) 1875--1899
 and arXiv:hep-ph/0411051

\bibitem{ze-bo-ha-ma-05c}
N. Zenine, S. Boukraa, S. Hassani and J.-M. Maillard,
{\em Square lattice Ising model susceptibility: connection matrices
         and singular behavior of $\chi^{(3)}$ and $\chi^{(4)}$},
 J. Phys. A: Math. Gen. {\bf 38 } (2005) 9439--9474
 and  arXiv:math-ph/0506065

\bibitem{Candelas}
P. Candelas, X. de la Ossa, P. Green and L. Parkes,
 {\em A pair of Calabi-Yau manifolds as an exactly
soluble superconformal theory}, Nucl. Phys. {\bf B359}, (1991), 21--74.

\bibitem{LianYau}
B.H. Lian and S-T. Yau,
 {\em Mirror Maps, Modular Relations and Hypergeometric Series II},
 Nuclear Phys. {\bf B 46}, Proceedings Suppl. Issues 1-3, (1996) 248--262
 and arXiv: hepth/950753v1 (1995)

\bibitem{Doran}
C. F. Doran,
 {\em Picard-Fuchs Uniformization and Modularity of the Mirror Maps},
 Comm. Math. Phys. {\bf 212}, 625-647, (2000). 

\bibitem{Doran2}
C. F. Doran,
 {\em Picard-Fuchs Uniformization: Modularity of the Mirror Map
 and Mirror-Moonshine},  CRM Proc. Lecture Notes, {\bf 24}, Amer. Math. Soc.
257--281, Providence
and arXiv:math/9812162v1,  (1998). 

\bibitem{Kratten}
C. Krattenthaler and T. Rivoal,
 {\em On the Integrability of the Taylor Coefficients
 of Mirror Maps},
 http://www-fourier.ujf-grenoble.fr/~rivoal, (2007). 

\bibitem{Atkin} H. H. Chan and M.-L. Lang, {\em Ramanujan's modular equations
    and Atkin-Lehner involutions}, Israel Journal of Mathematics, 
  {\bf 103}, (1998) pp. 1--16.

\bibitem{bo-bo-ha-ma-we-ze-09}
A. Bostan, S. Boukraa, S. Hassani, J.-M. Maillard, J.-A. Weil and N. Zenine,
{\em Globally nilpotent differential operators and the square Ising model},
J. Phys. A: Math. Theor. {\bf 42} (2009) 125206 (50pp) and arXiv:0812.4931

\bibitem{j} $j$-Function. http://mathworld.wolfram.com/j-Function.html

\bibitem{Ince} E. Ince,   {\em Ordinary Differential Equations}, 
New York: Dover, 1956, Dover Books in Mathematics

\bibitem{Vidunas} R. Vidunas, {\em Algebraic Transformations of Gauss
Hypergeometric Functions}, Funkcialaj Ekvacioj, {\bf 59}, (2009) 139-180.

\bibitem{Rota} K. Ebrahimi-Fard and L. Guo, 
{\em Rota-Baxter Algebras in Renormalization 
of Perturbative Quantum Field Theory},
	 Phys.Lett. {\bf B 281}, (1992), pp. 315--319 
         and arXiv:hep-th/0604116v2

\bibitem{Rota2} K. Ebrahimi-Fard, D. Manchon, F. Patras
{\em A noncommutative Bohnenblust-Spitzer identity for
 Rota-Baxter algebras solves Bogoliubov's recursion}, 
J. Noncommutative Geom. {\bf 3}, (2009), no.2, 181--222 and 
arXiv:0705.1265v2 [math.CO]

\bibitem{Siegel} C. L. Siegel, 
{\em Iteration of Analytic Functions}, 
Ann. Math. {\bf 43} 1942, 607--612

\bibitem{Siegel2} R. P\'erez-Marco, 
{\em S\'eminaire Bourbaki. Paris (1992)}, 
44\`eme ann\'ee 1991-92, {\bf 206}, Exp. no. 753, 4, 273-310,
 {\em Solution compl\`ete au probl\`eme 
de Siegel de lin\'earisation d'une application holomorphe
 au voisinage d'un point fixe (d'apr\`es Yoccoz)}

\bibitem{Almost} S. Boukraa, J-M. Maillard and G. Rollet, 
{\em Almost integrable mappings}, 
Int. Journ. of Modern Phys. {\bf B 8} 1994, 137--174

\bibitem{wu-mc-tr-ba-76}
T. T. Wu, B. M. McCoy, C. A. Tracy and E. Barouch, 
{\em Spin-spin correlation functions for the two-dimensional Ising model: 
Exact theory in the scaling region},
Phys. Rev. {\bf B 13} (1976) 316--374 

\bibitem{nickel-99} B. Nickel,
{\em On the singularity structure of the 2D Ising model susceptibility},
J. Phys. A {\bf 32} (1999), no. 21, 3889--3906.

\bibitem{nickel-00}
B. Nickel,
{\em Addendum to ``On the singularity structure
 of the 2D Ising model susceptibility''},      
J. Phys. A {\bf 33} (2000), no. 8, 1693--1711

\bibitem{growth} N. Abarenkova,  J-C. Angl\`es d'Auriac,
	S. Boukraa and J-M. Maillard,
 	{\em  Growth-complexity spectrum of some discrete dynamical systems},
        Physica {\bf D 130} (1999) pp. 27--42 and arXiv/chao-dyn/9807031

\bibitem{Functional} J-C. Angl\`es d'Auriac, S. Boukraa  and J-M. Maillard,
 	{\em  Functional relations in lattice statistical mechanics,
	 enumerative combinatorics and discrete dynamical systems}, 
       Annals of Combinatorics {\bf 3}, (1999) pp. 131--158

\bibitem{High}
A. Bostan, S. Boukraa, A.J. Guttmann, S. Hassani, I. Jensen, J.-M. Maillard, 
 and N. Zenine,
 {\em High order Fuchsian equations for the square Ising model:
 $\tilde{\chi}^{(5)}$},
J. Phys. A: Math. Theor. {\bf 42} (2009)  275209--275241
 and  arXiv:0904.1601v1 [math-ph]

\bibitem{bo-gu-ha-je-ma-ni-ze-08}
S. Boukraa, A.J. Guttmann, S. Hassani, I. Jensen, J.-M. Maillard, B. Nickel
 and N. Zenine,
 {\em Experimental mathematics on the magnetic susceptibility of
 the square lattice Ising model},
 J. Phys. A: Math. Theor. {\bf 41} (2008) 455202 (51pp)
 and  arXiv:0808.0763

\bibitem{ze-bo-ha-ma-05b}
N. Zenine, S. Boukraa, S. Hassani and J.-M. Maillard,
{\em Ising model susceptibility: Fuchsian differential 
equation for $\chi^{(4)}$
and its factorization properties},
   J. Phys. A: Math. Gen. {\bf 38} (2005) 4149--4173
 and  arXiv:cond-mat/0502155  

\bibitem{Holonomy} S. Boukraa, S. Hassani, J-M. Maillard,
 B. M. McCoy, W. Orrick, N. Zenine,
       	{\em Holonomy of the Ising model form factors},
	\newblock J. Phys. {\bf A 40}, 75-112 (2007), IOP Select and 
        \newblock http://arxiv.org/abs/math-ph/0609074  

\bibitem{higher3}
	  S. Boukraa, S. Hassani, J-M. Maillard, B. M. McCoy, N. Zenine,
	{\em The diagonal Ising susceptibility},
	\newblock  J. Phys. {\bf A 40}: Math. Theor. (2007) 8219-8236  
	\newblock and http://arxiv.org/pdf/math-ph/0703009 v2

\bibitem{maier-05}
R. Maier,
{\em On reducing the Heun equation to the hypergeometric equation},
J. Differential Equations {\bf 213} (2005), no. 1, 171--203.

\bibitem{Painleve} S. Boukraa, S. Hassani, J-M. Maillard,
 B. M. McCoy, J-A. Weil, N. Zenine,
	{\em Painlev{\'{e}} versus Fuchs},
	\newblock  J. Phys. {\bf A 39} (2006) 12245-12263
         Special issue for the centenary of 
         the publication of the Painlev\'e VI equation
        \newblock http://arxiv.org/pdf/math-ph/0602010 v3 

\bibitem{Fuchs}  S. Boukraa, S. Hassani, J-M. Maillard,
 B. M. McCoy, J-A. Weil, N. Zenine,
	{\em Fuchs versus Painlev{\'{e}}},
	\newblock J. Phys. {\bf A 40} (2007), 2583-2614
        \newblock Special issue for the SIDE VII conference (Melbourne)
	\newblock http://arxiv.org/pdf/math-ph/0701014 v2

\bibitem{Stiller}
P. F. Stiller,
 {\em Classical Automorphic Forms and Hypergeometric Functions},
 Journ. of Number Theory, {\bf 28}, no. 2,  219-232, (1988). 

\bibitem{Zudilin}
W. Zudilin,
 {\em The Hypergeometric Equation and Ramanujan Functions},
 The Ramanujan Journal, {\bf 7}, no. 4, 435-447, (2003). 

\bibitem{Andre} Y. Andr\'e,  
{\em G-functions and geometry}, Aspect of Mathematics E,
Num. 013, 
Vieweg Editor, (1989), ISSN: 0179-2156. 

\bibitem{Andre2} Y. Andr\'e and  F.  Baldassarri,
 {\em Geometric theory of $G$-functions},
  Arithmetic geometry (Cortona, 1994),  1--22, 
Sympos. Math., XXXVII, Cambridge Univ. Press,
Cambridge, (1997).

\bibitem{Andre3} Y. Andr\'e,  {\em  Arithmetic Gevrey 
series and transcendence. A survey},
 Journal de Th\'eorie des Nombres de Bordeaux, {\bf 15},
 (2003), no. 1, 1--10.

\bibitem{Andre4} Y. Andr\'e, {\em Sur la conjecture des $p$-courbures
 de Grothendieck-Katz et un probl\`eme de Dwork}, 
Geometric aspects of Dwork theory. Editors A. Adolphson,
 F. Baldassarri, P. Berthelot, N. Katz and F. Loeser,
Vol. {\bf I, II},  55--112, 
 Walter de Gruyter, Berlin, New-York, (1991).



\end{thebibliography}
\end{document}